\pdfoutput=1
\documentclass[3p,times]{article}

\usepackage{amssymb}
\usepackage{graphicx}
\usepackage[margin=1in]{geometry}

\newcommand{\beqn}{\begin{eqnarray}} 
\newcommand{\eeqn}{\end{eqnarray}} 
\newcommand{\halb}{\frac{1}{2}}

\newcommand{\no}{\nonumber}

\newcommand{\diag}{\mathrm{diag}}

\let\rho=\varrho
\newcommand{\hone}{{\mathbf 1}}
\newcommand{\hl}{{\mathbf l}}
\newcommand{\hr}{{\mathbf r}}
\newcommand{\sqrho}{\sqrt{\rho}}
\newcommand{\Sqrho}{\sqrt{\rho}}
\newcommand{\JJ}{J}

\begin{document}




\title{A conservative, skew-symmetric Finite Difference Scheme for the compressible Navier--Stokes Equations}

\author{Julius Reiss, J\"orn Sesterhenn\\{\small ISTA, Technische Universit\"at Berlin, Germany}}

\maketitle
\begin{abstract}
We present a fully conservative, skew-symmetric finite difference scheme on transformed grids. 
The skew-symmetry preserves the kinetic energy by first principles,  simultaneously avoiding a central instability mechanism
 and numerical damping.
In contrast to other skew-symmetric schemes no special averaging procedures are needed. 
Instead, the scheme builds purely on  point-wise operations and derivatives. 
Any explicit and central derivative can be used,  permitting high order and great freedom to optimize the scheme otherwise.
This also allows the simple adaption of existing finite difference schemes to improve their stability and damping properties. 
\end{abstract}


\section{Introduction}
Flows featuring   shocks  and acoustics are common in engineering applications. Examples are the sound generation of supersonic jets, noise
 generation in transonic flight or shock buffeting on an airfoil, where one possible mechanism is the interaction of sound waves
  from the trailing edge with the standing shock on the airfoil.

Simulations of such configurations are numerically challenging  as the correct and stable simulation of shocks and the faithful calculation of 
acoustics have different, and in standard approaches, contradictory requirements. Shocks, on the one hand, are described by the Rankine-Hugoniot conditions,
 which build directly on the conservation laws. Conservation of  mass,
  momentum and total energy is a prerequisite to guarantee a correct shock treatment. Such conserving schemes are usually finite volume (FV) schemes.  

Acoustics, on the other hand,  are propagating disturbances. They experience very low damping and travel long distances without noticeable energy loss in practice.
 When the amplitude is small the dispersion is also very small. To preserve these properties in a numerical simulation, schemes with very low (numerical)
   damping and low dispersion are needed, so that high-order, dispersion optimized finite difference (FD) schemes are mostly used. 

In contrast to finite volume, most finite difference schemes are only approximately conserving, becoming worse, where flow variables are
 rapidly changing. Thus, in shocks, where conservation is most desired they become unreliable. Indeed, it is well known that 
 FD schemes can totally fail to describe shocks. On the other hand FV usually utilise upwind schemes for stability reasons,
  leading to high artificial damping.  Even the optimized schemes hardly reach the quality of FD schemes for acoustics simulations.  
Thus a conservative scheme with low damping is desirable. 

To understand the key point of a low damping scheme consider the momentum equation for the $\alpha^{th}$ velocity component, ($\alpha,\beta\in 1,2,3$), ${\bf u}  = (u_1,u_2,u_3)^t $.    
$$
\partial_t \rho u_\alpha +\partial_{x_\beta} \left(\rho { u_\beta} u_\alpha\right) + \partial_{x_\alpha} p = \partial_{x_\beta} \tau_{\alpha\beta} .  \no
$$ 
As usual $p$ is the pressure and   $\tau_{\alpha\beta}= \mu(\partial_{x_\alpha}u_\beta + \partial_{x_\beta}u_\alpha) +(\mu_d -\mu2/3)  \delta_{\alpha\beta}\partial_{x_\gamma}u_\gamma$ the friction. Summing convention is assumed. 
From this the equation for the kinetic energy $E_{kin} =\rho {u_\alpha u_\alpha}/{2}$ is derived with the help of the mass 
conservation\footnote{This derivation will be shown in more detail further down.} as
$$
\partial_t \rho {u_\alpha u_\alpha}/{2} +\partial_{x_\beta}\left( \rho  u_\beta{ u_\alpha u_\alpha}/{2} \right) = -  { u_\alpha}\partial_{x_\alpha}  p + u_\alpha\partial_{x_\beta} \tau_{\alpha\beta}.  
$$
Only pressure work and friction changes the kinetic energy. The transport of the kinetic energy is in contrast strictly conservative. 
This physical property is easily destroyed in  numerical schemes.
 Up-winding  destroys the conservativity of the transport by introducing artificial damping; but even a central 
 derivative usually does not exactly preserve the kinetic energy. 
 Thus  a  transport term, which  conserves the kinetic energy is the key to an undamped simulation. 
 It should be pointed out, that an artificial damping does not destroy  the conservation of the {\it total energy} in FV, as by construction 
 the lost kinetic energy is balanced by an increased internal energy. 
 But the kinetic energy of, say, sound  waves is irreversible transformed to internal energy, thus to heat.    
\smallskip

The correct treatment of the kinetic energy is the focus of skew-symmetric schemes. 
The discretization of the non-linear transport term is chosen, so that the implied term 
for the equation of the kinetic energy is strictly conservative. 
This is achieved by formulating the transport term as a skew-symmetric operator, which implies 
the conservation of kinetic energy by first principles: 
the change of kinetic energy is calculated from a quadratic form of this transport term, 
 which is zero, as all quadratic forms of skew-symmetric matrices are zero.  
The  main challenge then is to preserve the conservation of the mass, momentum and total energy; 
especially the conservation of momentum is far from obvious and might be violated. 
To this end it has to be possible to rewrite the discrete skew-symmetric operator into a term with 
the telescoping sum property, typically to a discrete form of the divergence. 

The classical ansatz to insure this telescoping sum property of the skew-symmetric operator is to use skilful averaging of different variables. 
This was successfully applied by Morinishi in \cite{Morinishi1998} and became standard in this area. 
This procedure is by no means the only way to obtain skew-symmetry and the telescoping sum property;  
we find that no such averaging is necessary.  A straightforward and consistent discretization is sufficient. 
This not only leads to simpler expressions and to simple  proofs of all claimed properties, but it also allows to rewrite existing FD codes with minimal effort to a skew-symmetric form, by simply changing the spatial discretization, yielding good stability properties and low numerical damping. 
For strictly conservative schemes the time-stepping has to be changed accordingly; a dedicated paper about different time integration 
is submitted, \cite{Brouwer2013}. 
 
In contrast to the usual averaging procedure, the structure derived in this paper builds on matrices and  makes 
 little reference on the details of the stencil. Instead, abstract properties, namely the skew-symmetry and the telescoping sum property are assumed, 
 which are fulfilled by basically all central (and in computational space  equidistant)  derivatives. 
These properties permit one to choose a derivative to one's needs, be it any high order or wave-number optimized derivative. 
High order is obtained by using a standard (explicit) 
 high order derivative. 
  We make use of this freedom by choosing a derivative with the so called summation by parts property, \cite{Carpenter1994220}, 
  which allows clean and flexible boundary conditions without using ghost points.

For practical calculation curvilinear grids are essential. 
It is understood, \cite{VerstappenVeldman2003}, that a grid transformation to a computational space is a suitable way,
 to preserve the correct structure on curvilinear grids.
  To our knowledge the first to obtain this for compressible flows was Kok, \cite{Kok2009};  
the authors presented a similar procedure for our FD scheme in \cite{Reiss2010}. 
During submission we became aware of \cite{Morinishi2013}, which builds on similar ideas and extends it even to moving grids.  

A third way to derive skew symmetric-schemes is the Galerkin ansatz. 
Products in the function space can be approximated leading to a numerically effective scheme, see e.g. Gassner \cite{Gassner2013}. 
The approximation leads a skew-symmetric  scheme for the Burgers equation with similar point-wise products and derivatives as derived in this paper;
 the summation by parts property which  is a choice in this paper occurs naturally in the work by Gassner.
\smallskip 

This paper is organized as follows: first we introduce the rewriting and our way of spatial discretization of the Navier-Stokes
 equations in one dimension in  section \ref{OneDimensionSection}. Then we derive the three-dimensional equations, 
  introduce  curvilinear grids and discretize in section \ref{secNS3d}. A time discretization which generalizes 
  the scheme of Morinishi  \cite{Morinishi2010} and Subbareddy et al. \cite{Subbareddy2009} is derived in section \ref{time}.   
Boundaries are discussed in section \ref{boundaries}. We close with some numerical examples in section \ref{numericalExamples}. 
In the appendix we first compare our scheme with the standard, averaging approach.  
Finally we show by construction, that the FD scheme implies local and consistent fluxes.

\section{The Navier-Stokes Equations in one Dimension}
\label{OneDimensionSection} 
	
The Navier-Stokes equations in one dimension are given by
\beqn
\partial_t \rho  + \partial_x(\rho u)&=&0 \label{NSMasse} \\
\partial_t (\rho u)   + \partial_x(\rho u^2)+ \partial_x p  \label{NSImpuls} &=& \partial_x \tau  \\
\partial_t (\rho e + \rho u^2/2) + \partial_x \left(\rho u (e + p/\rho+ u^2/2)  \right) &=& \partial_x u \tau + \partial_x \phi .   
\eeqn
They describe mass, momentum and energy conservation. 
In the following  the internal energy of the ideal gas  $e=(p/\rho) /(\gamma-1)$ is assumed with the constant adiabatic index $\gamma$. 
Perfect gas is assumed in many simulations. However, we  emphasize that the following steps concern only the kinetic energy, 
 so that  any other equation of state could be used. The friction in one dimension is $\tau = \mu \partial_x u$ with $\mu$ the 
 viscosity and  $\phi = \lambda \partial_x T $ is the heat flux, with the heat conductivity $\lambda$ .       

The transport term in the momentum equation (\ref{NSImpuls}) given in divergence form, can be rewritten to convective form with help of 
the mass conservation as 
$$
\partial_t (\rho u)   + \partial_x(\rho u^2) = \rho \partial_t  u   + \rho u \partial_x u.$$
Adding the divergence and convective form of the momentum equation we obtain the skew-symmetric form.  The Navier-Stokes equations become  
\beqn
\partial_t \rho  + \partial_x(\rho u)&=&0 \\
\halb \left(\partial_t\rho\cdot + \rho\partial_t\cdot\right)u
     +\halb \left( \partial_x u\rho\cdot + u\rho\partial_x \cdot\right)u+\partial_x p&=& \partial_x \tau \label{NSSkewImpuls}\\
\frac {1}{\gamma-1}\partial_t p   + \frac {\gamma}{\gamma-1}  \partial_x \left( u p   \right)-u \partial_x p&=&
-u\partial_x  \tau  +\partial_x u \tau + \partial_x \phi , 
\eeqn
where it is understood that the space and time derivatives in the first two terms 
of (\ref{NSSkewImpuls}) act also on $u$ right of the parentheses. For clarity this is explicitly marked by a dot ,,$\cdot$''. 
The kinetic energy was split off from the energy equation with the help of the momentum equation by use of the product rule 
\beqn
&&\partial_t (  \rho u^2/2) + \partial_x \left(\rho u ( u^2/2)  \right) \no\\
&=&  \halb u ( \partial_t \rho \cdot  +  \rho  \partial_t  \cdot ) u + \halb u (\partial_x (\rho u) \cdot +  (\rho u) \partial_x \cdot ) u\no\\
& = & - u \partial_x p  + u\partial_x \tau,
\eeqn 
leaving just the pressure work $-u\partial_x p$ and a friction term. 
In the second line the skew-symmetric form of the momentum transport term appears, 
which underlines its close connection to the kinetic energy.   

Straightforward discretization leads to 
\beqn
  \partial_t  \rho +       B^{ u}  \rho  &=&0 \label{mass1Ddisc}\\[0pt]
\halb (\partial_t   \rho \cdot+  \rho \partial_t \cdot)   u +  \halb D^{  u\rho }  u 
 +   D_x  p &=&  D_x \tau \label{mom1Ddisc}\\  
 \frac{1}{\gamma-1} \partial_t  p + \frac {\gamma}{\gamma-1}    B^{ u}  p -   C^{ u}   p &=& - UD_x\tau + D U \tau + D_x \phi  \label{en1Ddisc},
\eeqn
where we have introduced the matrices 
\beqn
B^{ u} = D_x U ,\quad  C^{u} = U D_x \quad\mathrm{and\quad}
D^{u \rho} = (D_x U R + R UD_x) 
\eeqn
with skew-symmetric derivatives\footnote{We hereby assume periodic derivatives. Boundaries are discussed further down.} $D_x=-D_x^T$. 
The capital letters stand for  $U = \diag(u) , R = \diag(\rho)$, mimicking  the point-wise product.
 Beside the skew-symmetry we also need the telescoping sum property, or method of 
 differences, $\sum_i (D_x)_{ij}=0$. 
 As we  assume equidistant spacing in calculation space this is trivially fulfilled. 
 We abbreviate this sum with ${\bf 1}^T=(1,1,1,\dots) $, so that ${\bf 1}^T D_x =  0$ and ${\bf 1}^T U = u^T $.  
The product terms of $u$ and $\rho$ in the time derivative in the momentum equation are assumed to be point-wise in general, to have clearer notation further down.   

First note that 
\beqn
D^{u \rho} = - (D^{u \rho})^T \label{dru1D}
\eeqn
 is skew-symmetric by construction. This name giving property leads right away to the conservation of the kinetic energy,
since quadratic forms of skew-symmetric matrices $u^T D^{u \rho} u = (u^T D^{u \rho} u)^T  =- u^T D^{u \rho} u = 0 $ vanish. 
Indeed, the change of the kinetic energy can be calculated as 
\beqn
0&=&\phantom{-}
   \halb\int u^T\left(\partial_t \rho \cdot + \rho  \partial_t \cdot \right) u dt \phantom{+[u^T \rho u]_{t_0}^{t_1}\cdot} + \halb \int u^T D^{u \rho } u dt 
+ \int u^T (D_x p -D_x\tau)   dt\no\\
&=&-\halb\int u^T\left(\partial_t \rho \cdot+ \rho  \partial_t\cdot \right) u dt+ [u^T \rho u]_{t_0}^{t_1} -\halb  \int u^T D^{u \rho } u dt 
+ \int u^T (D_x p -D_x\tau)   dt.   \no
\eeqn 
Adding the two lines $ (1/2)[u^T \rho u]_{t_1}^{t_0}  = \int u^T D_x p +u^T D_x\tau  dt  $ we find that the change of energy is given by
 the pressure work and friction alone, as stated in the introduction. 
 The {\it transport} of kinetic energy by the momentum equation is conservative, as intended.  

 We now show further, that this skew-symmetry and 
 the easy-to-check identities 
\beqn
\hone^T B^{ u} &=& 0 \label{oneBu} \\
\hone^T  D^{  u\rho }  u & =&  -u^T  B^{ u} \rho \label{DruBu} \\
\hone^T C^u p  &=& u^T D_x p   \label{cup},
\eeqn
are sufficient to prove the  conservations of mass, momentum and energy. \\ 
Summing the mass equation ($ \hone^T (\ref{mass1Ddisc} )$)  we find  with eqn. (\ref{oneBu})  the mass conservation  
\beqn
\partial_t \hone^T \rho = -\hone^T B^{ u}  \rho = 0.
\eeqn
The momentum conservation is found from the combination of the mass and momentum equation 
$ (\ref{mom1Ddisc})  + \halb u^T (\ref{mass1Ddisc}) $ with eqn. (\ref{DruBu}) to be  
\beqn
 \partial_t (\rho^T u )&=& \halb u^T \partial_t  \rho  + \hone^T \halb (\partial_t   \rho \cdot+  \rho \partial_t \cdot)   u  \no\\
&=& -\halb \left( u^T  B^{ u}  \rho + \hone^T  D^{  u\rho } \right) - \hone^T D_x p
 = 0  .
 \eeqn
The total energy is conserved as the combination of momentum and energy equation $  \hone^T (\ref{en1Ddisc})+ u^T (\ref{mom1Ddisc})   $ is with the help of (\ref{dru1D} and  \ref{oneBu})  found to be 
\beqn
\partial_t \left(  \frac{\hone^T p}{\gamma-1} + \frac{\rho^T u^2}{2}\right)
&=&  \partial_t   \frac{\hone^T p}{\gamma-1} +  \halb u^T  (\partial_t   \rho \cdot+  \rho \partial_t \cdot)   u\no\\
&=& -\halb u^T D^{u \rho} u  - u^T D_x p +  u^T D_x p \no\\ 
&=&0.  
\eeqn
In the last step the quadratic form $u^T D^{u \rho} u$ vanishes due to the skew-symmetry of $D^{u \rho}$. 
Note that the conservation of momentum and energy is not a direct consequence of one equation as in finite volume schemes, but due to the 
consistent  discretization of the set of equations.
In the next section the multi dimensional analogues  of the last expressions are derived. 
	
Note that only little assumptions about the structure of the derivative were made. Indeed, all explicit, central derivatives can be used. 
Compact derivatives can also be used for periodic grids; for non-periodic grids the implied derivative matrix will typically 
break the strict skew-symmetry also far away from the boundary.
  
Note further that in contrast to other skew symmetric schemes no averaging was used. A detailed discussion of this difference is
given in \ref{appendixCompare}.

\section{More Dimensions and Grid transformations}
\label{secNS3d}

To preserve the skew-symmetry and conservation it is crucial to keep the structure of the one dimensional operators  in more dimensions. 
On Euclidean grids this is straightforward, but it poses an additional difficulty on curvilinear grids, which are important for
 any realistic geometry. Namely the skew-symmetry can  easily be broken on transformed, especially on {\it non-orthogonal} grids. 
Ducros et. al. \cite{DucrosEtAl2000} use two different approaches for transformed grids, first they use  the chain rule directly,
 and secondly an interpolation method,  but in general they have to either sacrifice the conservation or the skew-symmetry. 

It was also already noted by Verstappen et. al. \cite{VerstappenVeldman2003} that the use of transformations, instead 
of modifying the derivative stencil, is advantageous. The first working approach, to our knowledge, was presented by Kok in 2009, \cite{Kok2009}, 
in FV context. For FD we presented our  approach on the STAB 2010 conference \cite{Reiss2010}, utilizing similar ideas. We use  
this approach in the  following part, as it allows to obtain equations structurally as simple as in one dimension.
During submission we became aware of \cite{Morinishi2013}, which is similar to Kok's and our approach. 
Seemingly, grid transformations  and local bases are the correct way to preserve skew-symmetry and conservation for distorted grids. 

\subsection{Analytical equations in 3D}
The multi-dimensional Navier-Stokes Equations are
\beqn
\partial_t \rho + \partial_{x_\beta} \rho  u_\beta &=& 0 \label{mass3D}\\
\partial_t \rho u_\alpha + \partial_{x_\beta} (\rho {u_\beta} u_\alpha ) + \partial_{x_\alpha}  p  &=& \partial_{x_\beta} \tau_{\alpha\beta} 
\label{mom3D} \\
 \partial_t \left(\rho\left[ e + \frac{u_\alpha u_\alpha}2\right] \right)  
+ \partial_{x_\beta} \left( \rho u_\beta \left[ e +\frac{u_\alpha u_\alpha}2 +\frac p \rho\right] \right)    & =&  
\partial_{x_\alpha} u_\beta \tau_{\alpha \beta}  + \partial_{x_\alpha} \phi_\alpha.
\label{energy3D}
\eeqn

The Greek letters mark spatial direction $  \beta = 1,2, 3$, and summing convention is assumed. 
We use the shorthand notation $ \partial_{x_\beta} = \frac \partial {\partial_{x_\beta} }$. The dissipative terms were defined 
in the introduction. 

As in one dimension, the convective term in the momentum equation (\ref{mom3D}) can be rewritten from divergence (D) 
to convective (C) form: 
\beqn
\partial_t \rho u_\alpha + \partial_{x_\beta} (\rho {u_\beta} u_\alpha )  
=
\rho\partial_t  u_\alpha + \rho {u_\beta}\partial_{x_\beta} (  u_\alpha ). 
\eeqn
Taking again as in one dimension $(D+C)/2$ we arrive at the skew-symmetric form of the momentum equation 
\beqn
\halb \left(\partial_t\rho\cdot + \rho\partial_t\cdot\right)u_\alpha
     +\halb \left( \partial_{x_\beta} u_\beta\rho\cdot + u_\beta\rho\partial_{x_\beta} \cdot\right)u_\alpha
     + \partial_{x_\alpha}  p  &=& \partial_{x_\beta} \tau_{\alpha\beta}
\eeqn
 The change of kinetic energy can be expressed, as in one dimension, as  
\beqn
&&\partial_t \left(  \rho \frac{u_\alpha u_\alpha}2\right) + \partial_{x_\beta} \left(\rho u_\beta  \frac{u_\alpha u_\alpha}2  \right) \no\\
&=&  \halb u_\alpha ( \partial_t \rho \cdot                  +  \rho  \partial_t  \cdot                 ) u_\alpha +
     \halb u_\alpha (\partial_{x_\beta} \rho u_\beta \cdot +  \rho u_\beta \partial_{x_\beta} \cdot ) u_\alpha \no\\
& = & - u_\alpha \partial_{x_\alpha} p  + u_\alpha \partial_{x_\alpha} \tau_{\alpha\beta},
\eeqn 
so that the energy equation becomes 
\beqn
 \partial_t \left(\rho e  \right)  
+ \partial_{x_\beta} \left(  u_\beta ( \rho e + p ) \right)  
- u_\alpha \partial_{x_\alpha} p
  & =& 
- u_\alpha \partial_{x_\alpha} \tau_{\alpha\beta}
+\partial_{x_\alpha} u_\beta \tau_{\alpha \beta}  + \partial_{x_\alpha} \phi_\alpha.
\eeqn  
In this paper we will always assume the ideal gas law  $ \rho e = p/(\gamma-1)$,  with a constant adiabatic index $\gamma$.   
 
The set of equation therefore is:
\beqn
\partial_t \rho + \partial_{x_\beta} \rho  u_\beta &=& 0  \\ 
\halb \left(\partial_t\rho\cdot + \rho\partial_t\cdot\right)u_\alpha
     +\halb \left( \partial_{x_\beta} u_\beta\rho\cdot + u_\beta\rho\partial_{x_\beta} \cdot\right)u_\alpha
     + \partial_{x_\alpha}  p  &=& \partial_{x_\beta} \tau_{\alpha\beta} \label{mom3Dskew}\\ 
\frac 1 {\gamma -1} \partial_t  p    
+ \frac \gamma  {\gamma -1} \partial_{x_\beta} \left(  u_\beta   p  \right)  
- u_\alpha \partial_{x_\alpha} p
  & =& 
- u_\alpha \partial_{x_\beta} \tau_{\alpha\beta}
+\partial_{x_\beta} u_\alpha \tau_{\alpha \beta}  + \partial_{x_\alpha} \phi_\alpha 
\eeqn
 The structure is essentially the same as in one dimension.  Therefore all discussed properties are the same.  
  
\subsection{Transformed grids}
Similar to a previous work of the authors \cite{Reiss2010}, we now introduce grid transformations.

Grid transformations are mappings between the physical coordinates $x_\alpha=(x,y,z)= (x_1,x_2,x_3)$ and a computational
 space  $(\xi,\eta,\zeta) = (\xi^1,\xi^2,\xi^3) $, the $x_\alpha$ are understood as functions of $\xi^\beta$ and vice versa. 
  The computational space is a unit cube in which the discretization is 
 equidistant $(\xi^{\alpha}_i = i\Delta \xi^{\alpha}  )$. 
 
 As can be found in    \cite{Thompson1985},
  the divergence or gradient can be expressed in two forms,  named {\it conservative} (\ref{nabCons})  and {\it non-conservative} (\ref{nabNonCons}) ,    
\beqn
&&\frac{  \partial u_\beta}
{\partial x_\beta} = \nabla {\bf u } \no\\
&=& \frac 1 {J} \sum_{\gamma_i,\mathrm{cy}} \partial_{\xi^{\gamma_1}} ({\bf e}_{\gamma_2}\times {\bf e}_{\gamma_3}) {\bf u }  \label{nabCons}\\
&=& \frac 1 {J} \sum_{\gamma_i,\mathrm{cy}}  ({\bf e}_{\gamma_2}\times {\bf e}_{\gamma_3}) \partial_{\xi^{\gamma_1}} {\bf u }.\label{nabNonCons} 
\eeqn
where the indices $\gamma_1,\gamma_2,\gamma_3$ are assumed to be cyclic, meaning ((1,2,3), (2,3,1) or (3,1,2)), implying  $\gamma_2 = \gamma_2(\gamma_1)$, $\gamma_3 = \gamma_3(\gamma_1)$. $J=({\bf e}_1\times {\bf e}_2)\cdot {\bf e}_3$ is the Jacobian
of the grid transformation. It will reappear  in a natural way as an integration weight in the internal expressions of the conserved quantities. 
The local base vectors are defined as
\beqn
{\bf e}_\alpha= \partial_{\xi^\alpha} {\bf r}  \label{baseVec}, 
\eeqn  
with $r= (x,y,z)^T$.

Our starting point will is the skew-symmetry of the transport term in the momentum equation (\ref{mom3Dskew}). 
If we use (\ref{nabCons}) for the first, the divergence term, and (\ref{nabNonCons}) for the second, the gradient 
term,  the symmetry  of the expression is preserved, which will lead to the correct skew-symmetry of the operator in discrete sense:
\beqn
&&      \left( \partial_{x_\beta} u_\beta\rho\cdot + u_\beta\rho\partial_{x_\beta} \cdot\right)u_\alpha\no\\
&=& \frac 1 J  \sum_{\gamma_i,\mathrm{cy}} \left(
\partial_{\xi^{\gamma_1}} ({\bf e}_{\gamma_2}\times {\bf e}_{\gamma_3}) {\bf u }\rho  \cdot
+ 
 ({\bf e}_{\gamma_2}\times {\bf e}_{\gamma_3}) {\bf u }\rho \partial_{\xi^{\gamma_1}} \cdot
\right)   
u_\alpha\no\\
&=& 
\frac 1 J  \sum_{\gamma} \left(
\partial_{\xi^{\gamma}} \tilde u_\gamma \rho  \cdot
+ 
 \tilde u_\gamma \rho \partial_{\xi^{\gamma_1}} \cdot
\right)   
u_\alpha\label{skewTrans3d}
\eeqn
where we have introduced the effective velocity components 
\beqn
\tilde u_{\gamma_1} = ({\bf e}_{\gamma_2}\times {\bf e}_{\gamma_3}) {\bf u } \qquad\mathrm{with\; \gamma_i\; cyclic}.
\label{uTilde}
\eeqn
Obviously every operator term in the sum of (\ref{skewTrans3d}) has the correct skew-symmetry.  
The other derivative operators have to be chosen consistently. 
One possible choice is to use the conservative form for all other Euler terms.  We thus arrive at 
\beqn
J \partial_t \rho + \partial_{\xi^\beta} \tilde u_\beta \rho &=& 0 \\
J \halb \left(\partial_t\rho\cdot + \rho\partial_t\cdot\right)u_\alpha
     +\halb \left( \partial_{\xi^\beta}\tilde u_\beta\rho\cdot + \tilde u_\beta\rho\partial_{\xi^\beta} \cdot\right)u_\alpha
     + J\partial_{x_\alpha}  p  &=& \partial_{\xi_\beta} \tilde \tau_{\alpha\beta}\\
J \frac 1 {\gamma -1} \partial_t  p    
+ \frac \gamma  {\gamma -1} \partial_{\xi^\beta} \left( \tilde  u_\beta   p  \right)  
- J u_\alpha \partial_{x_\alpha} p
  & =& 
- u_\alpha \partial_{\xi^\alpha} \tilde \tau_{\alpha\beta}
+\partial_{\xi^\alpha} u_\beta \tilde \tau_{\alpha \beta}  + \partial_{\xi^\alpha} \tilde\phi_\alpha. 
\eeqn
Note the remaining $x_\alpha$ derivatives in the gradient of the pressure, which are now understood as the components of 
\beqn
J\partial_{x_\alpha}  p
= 
J (\nabla p)_\alpha = \left( 
\sum_{\gamma_i,cy} \partial_{\xi^{\gamma_1}} ({\bf e}_{\gamma_2}\times {\bf e}_{\gamma_3}) 
p \right)_\alpha. \label{Dx_curv}  
\eeqn   
The dissipative terms are 
discretized by using a non-conservative form for the inner derivative and a conservative form for the outer derivatives.
The reason is twofold: first this procedure leads to symmetric operators  
for this terms also on distorted grids, which is in line with its analytical properties. 
The strict destruction of kinetic energy is a direct consequence\footnote{These derivatives do not need to be 
skew-symmetric, but can have an unsymmetrical stencil. It it advisable, that the inner and outer derivative have $D_{in}= -D_{out}^T$ leading to a symmetric dissipative term. }. 
Secondly, the total number of derivatives is not increased compared with Cartesian grids, as this allows to collect the inner-derivative terms. 
The heat flow is  
\beqn
\tilde \phi_\alpha  = m_{\alpha,\beta} \frac \lambda J m_{\beta,\xi^\gamma}  \partial_{\xi^\gamma} T,
\eeqn     
where we have used the abbreviation
\beqn
m_{\alpha,\gamma_1} = ({\bf e}_{\gamma_2}\times {\bf e}_{\gamma_3})_\alpha \qquad \mathrm{\gamma_i\;  cyclic}
\eeqn
where the subscript is the  $\alpha^{th}$ entry of the vector. The stresses are 
\beqn
\tilde \tau_{\alpha\beta} = m_{\beta, \gamma}  \tau_{\alpha\gamma}  ,
\eeqn   
where $\tau$ can calculated on the curvilinear grid as
\beqn
 \tau_{\alpha\beta} = \frac \mu J  
 \left(
 m_{\beta\gamma} \partial_{\xi^\gamma} u_{\alpha}   
 +
 m_{\alpha\gamma} \partial_{\xi^\gamma} u_{\beta}   
  \right)  
  + \left(\mu_d-\frac 2 3 \mu \right)
  \frac 1 J \partial_{\xi^\gamma} \tilde u_\gamma  .
\eeqn   

We find that the number of derivatives  on curvilinear grids is only slightly increased by the pressure gradient term. 
Usually the number of derivatives is a good indicator of the  performance of a FD code.

\subsection{Discrete equations in 3D} 
The discretization is again done by replacing all products simply by point-wise products and derivatives by
 derivative  matrices. The derivatives are simple one dimensional derivatives in any direction in calculation space.  
\beqn
J  \partial_t  \rho +       B^{ u}  \rho  &=&0 \label{mass2Ddisc}\\[0pt]
J \halb (\partial_t   \rho \cdot+  \rho \partial_t \cdot)   u_\alpha +  \halb D^{  u\rho }  u_\alpha 
 +   \bar D_{x_\alpha}  p &=& D_{\xi^\beta} \tilde \tau_{\alpha\beta}  \label{momUalphadisc}\\  
 J\frac{1}{\gamma-1} \partial_t  p + \frac {\gamma}{\gamma-1}    B^{ u}  p -   C^{ u}   p &=& 
 - u_\alpha D_{\xi^\alpha} \tilde \tau_{\alpha\beta}
+D_{\xi^\alpha} u_\beta \tilde \tau_{\alpha \beta}  + D_{\xi^\alpha} \tilde\phi_\alpha 
  \label{en2Ddisc} .
\eeqn
with 
\beqn
B^{ {\bf u}} &=& D_{\xi^\beta} \tilde  U_\beta      \label{Bu3d} \\
D^{{\bf u} \rho} &=& (D_{\xi^\beta}  \tilde  U_\beta R + R\tilde  U_\beta D_{\xi^\beta} )   \label{2dDru}\\
C^{\bf u} &= &U_\alpha \bar D_{x_\alpha}  .
\eeqn
Note that we still assume summing convention. Note also that the momentum equation still describe the change of $u_\alpha$, and 
not of the  effective velocities  $\tilde u_\alpha$. Also the term $C^{\bf u} $ is still formulated in the Cartesian velocity components. 
The bar in the derivative $\bar D_{x_\alpha}$ marks the extra factor $J$, see (\ref{Dx_curv}).    

The transport term is skew-symmetric, as it is skew-symmetric in every direction,
\beqn
D^{{\bf u} \rho}& =& - (D^{{\bf u} \rho})^T \label{DruDruT}
\eeqn
 For this expressions the multi-dimensional analogues of (\ref{oneBu}-\ref{cup}) hold, namely 
\beqn
\hone^T B^{\bf u } &=&0 \label{Bu0}\\
\hone^T  D^{  {\bf u}\rho }  u_\alpha  & =&  - u_\alpha^T  B^{ {\bf u} } \rho \label{DruBu0}\\
\hone^T C^{\bf u} p  &=& u^T_\alpha D_{x_\alpha}  p  .    \label{CupDD}
\eeqn
 
The conservation is thus easy to prove along the one dimensional proof;  the kinetic energy is 
described by all three momentum equations. 

Again,  we did not make any further assumptions about the derivative in computational space. 
Any central, one dimensional derivatives in each directions can be used, i.e. a derivative for which 
 $\hone^T D $ and $D = -D^T$ holds.  
 
 The equations become very simple and compact in two dimensions.  
Assuming that the grid transformation is $x=x(\xi,\eta)$, $y=y(\xi,\eta)$ and $z = \zeta$, and further, that  the solution does not depend on 
$\zeta$, the effective velocities reduce to $\tilde u_1 \equiv \tilde u = u y_\eta -v y_\xi  $ and 
$\tilde u_2 \equiv \tilde v =  -u x_\eta + v x_\xi  $ and the Jacobian becomes $J = x_\xi y_\eta- x_\eta  y_\xi$. 
As $u_3 \equiv 0$ for all times the connected momentum equation can be dropped.  
The pressure gradient becomes 
$ \bar D_x p =  D_\xi(y_\eta p ) - D_\eta(y_\xi p )$ and           
$ \bar D_y p = -D_\xi(x_\eta  p ) + D_\eta(x_\xi p )$. 
     
\section{Time discretization} 
\label{time}

Similar to the spatial  discretization, the time discretization has to preserve  the conservation of 
kinetic energy by the momentum transport and conserved quantities mass, momentum and energy. 
The unusual time derivative operator in (\ref{momUalphadisc}) poses an additional  difficulty. 
Two strategies are at hand; either a multi-step method can be constructed or the time derivative operator 
can be rewritten as in  \cite{Morinishi2007}. 

Central (in time) multi-step methods lead to Leap-Frog-like methods. 
They are stable as long as no friction is involved. We used them successfully in 
\cite{Reiss2010} for our skew-symmetric scheme. Friction terms can also be included easily, by using a mixed Euler-Leap-Frog method. A peculiar kind of conservation can be 
derived, where the conserved quantities are build from two successive time steps. Beside some practical difficulties, like conservative filtering,
this does not pose a norm in the mathematical sense and is thus not followed. 

Instead, we use Morinishi's rewriting of the time derivative operator, to  find a  time integration scheme, which is a generalization of the scheme which was independently  derived by Subbareddy et al. \cite{Subbareddy2009} and by Morinishi \cite{Morinishi2010}. 
While the derivation in the first paper is motivated by Roe's Riemann solver, 
the derivation of Morinishi builds on a skew-symmetric time derivative on spatio-temporal staggered  grids. 
Here we show, that after using the rewriting of the time derivative proposed  by Morinishi, an implicit midpoint rule can be modified to give a time integration  which generalizes the before mentioned results. The generalization to higher order is discussed in a parallel publication \cite{Brouwer2013}.

Morinishi's  rewriting \cite{Morinishi2007}  transforms the time derivative in the momentum equations (\ref{mom3D}) or (\ref{momUalphadisc}) to
\beqn
\halb \left( \partial_t \rho\cdot  + \rho \partial_t \cdot \right)  u_\alpha 
= \sqrt{\rho} \partial_t   (\sqrt{\rho}  u_\alpha ).\no
\eeqn   
The square-root of the mass is well defined as $\rho\ge 0$ holds. 
The unusual appearance of $\sqrho$ instead of $\rho$ in the momentum equation can be understood when multiplying it with $u_\alpha$
 to derive  the kinetic energy conservation, i.e.  
\beqn
 u_\alpha \sqrt{\rho} \partial_t   \sqrt{\rho}  u_\alpha  = \halb   \partial_t  (\sqrt{\rho}  u_\alpha )^2 =  \halb   \partial_t \rho  u_\alpha^2
\label{ptRule}.
\eeqn 
 It is thus  a quadratic splitting of the kinetic energy. 

 It will prove helpful to rewrite the mass equation in terms of $\sqrt{\rho}$ by inserting $ \rho = \sqrt{\rho}^2$.  All remaining $\rho$ will be understood as $\rho\equiv \sqrt{\rho}^2  $.
Altogether we arrive at  
\beqn
\JJ\Sqrho \partial_t  \sqrt{\rho} +    \halb    B^{\bf u}  \rho  &=&0 \label{moriMass}\\[0pt]
 \JJ\Sqrho \partial_t  \left( \sqrt{\rho}  u_\alpha\right) +  \halb D^{  {\bf u}\rho }  u_\alpha 
 +   D_{x_\alpha}  p &=& F^{u_\alpha} \label{moriMom} \\  
 \frac{\JJ}{\gamma-1} \partial_t  p + \frac {\gamma}{\gamma-1}    B^{ \bf u}  p -   C^{\bf u}   p &=& F^{e} \label{moriEn}, 
\eeqn
where we abbreviated the dissipative  terms in the momentum and energy equations as $F^{u_\alpha}$ and $F^{e} $.      

The conservations properties of the time derivative terms are easily checked analytically.
  We do this as a guideline for the discretization. 
We only discuss the time derivatives, the spatial part is analogous to what was discussed before.  
For the mass we need
\beqn
\sqrho \partial_t \sqrho  = \halb \partial_t\rho,\label{sqrohtRule} 
\eeqn
 therefore the time integration has to  conserve quadratic forms.
The momentum conservation is found by combining $(\ref{moriMom})+ u_\alpha \cdot(\ref{moriMass})$, as  
\beqn
\sqrt{\rho} \partial_t  \left( \sqrt{\rho}  u_\alpha\right) + u_\alpha \sqrt{\rho} \partial_t  \sqrt{\rho} = 
\partial_t   \sqrt{\rho}\sqrt{\rho}  u_\alpha\label{utRule} 
\eeqn 
and the conservation of the kinetic energy by $u_\alpha \cdot(\ref{moriMom}) $, again demands the conservation of quadratic forms (\ref{ptRule}), which leads right away to the conservation of the full energy. 
\medskip

The three statements (\ref{ptRule}), (\ref{utRule}) and (\ref{sqrohtRule}) have to hold also in the time discrete case. 
For (\ref{sqrohtRule}) this  is easily accomplish  by the implicit midpoint rule, as it preserves quadratic quantities. 
If we insert this approach in the Navier Stokes equations (\ref{moriMass}--\ref{moriEn}) and carefully check which terms have to agree we arrive 
at the fully discrete equations
 \beqn
\JJ\sqrho^{n+1/2}  \frac{\left(\sqrho^{n+1} - \sqrho^{n}  \right)}{\Delta t} +    \halb    B^{{\bf u }^{n+a} }  \rho^{n+b}  &=&0 
\label{moriMassDisc}\\[0pt]
\JJ \sqrho^{n+1/2}  \frac{\left( \sqrho u_\alpha\right)^{n+1} - \left(\sqrho u_\alpha\right)^{n} }{\Delta t} 
+  \halb D^{  {\bf u}^{n+a} \rho^{n+b} }  u_\alpha^{n+1/2}
 +   D_{x_\alpha}  p^{n+c} &=&  F^{u_\alpha,{n+f}} \label{moriMomDisc} \\  
 \JJ\frac{1}{\gamma-1} \frac{ ( p^{n+1} - p^{n+1})}{\Delta t} 
+ \frac {\gamma}{\gamma-1}    B^{{ \bf u}^{n+d}}  p^{n+e} 
-   C^{{\bf u}^{n+1/2} }  p^{n+c} &=& F^{e,{n+f}} \label{moriEnDisc}, 
\eeqn
  and $\left(\sqrho u_\alpha\right)^{n} =\sqrho^{n} u_\alpha^{n} $. 
 We still have to define the time averages. 
If we set  $\sqrho^{n+1/2}= (\sqrho^{n+1} - \sqrho^{n})/2$,
 the third binomial formula can be applied to the first term of  (\ref{moriMassDisc}) to reproduce (\ref{sqrohtRule})
\beqn
\sqrho^{n+1/2} (\sqrho^{n+1} - \sqrho^{n}) = (\sqrho^{n+1})^2 - (\sqrho^{n})^2 . \label{sqrBinom} 
\eeqn 
To use the same procedure  for the kinetic energy as (\ref{ptRule}), we consider $ u_\alpha^{n+1/2} \cdot (\ref{moriMomDisc})$ and thus  would like to have 
\beqn
&&u^{n+1/2}_\alpha \sqrho^{n+1/2}  \left( (\sqrho  u_\alpha)^{n+1} - (\sqrho u_\alpha)^{n} \right)\\
&\equiv&\halb \left( (\sqrho u_\alpha)^{n+1} + (\sqrho u_\alpha)^{n} \right) \left( (\sqrho u_\alpha)^{n+1} - (\sqrho u_\alpha)^{n} \right)\\
&=&  \halb\left( (\sqrho u_\alpha)^{n+1})^2 - ((\sqrho u_\alpha)^{n})^2 \right)
\eeqn
leading to the {\it definition} of 
\beqn
u^{n+1/2}_\alpha = \frac {\left( \sqrho^{n+1} u_\alpha^{n+1} + \sqrho^{n} u_\alpha^{n} \right)}{2 \sqrho^{n+1/2} }.
\eeqn
This  is exactly what was found by Morinishi and Subbareddy et al.. 
The momentum conservation following  (\ref{utRule}) is then easy to see, by  $(\ref{moriMomDisc})+ u_\alpha \cdot(\ref{moriMassDisc})$,  as 
\beqn
&&u^{n+1/2}_\alpha \sqrho^{n+1/2} \left(\sqrho^{n+1} 
- \sqrho^{n}  \right)+\sqrho^{n+1/2}  \left( (\sqrho u_\alpha)^{n+1} - (\sqrho u_\alpha)^{n} \right)\no\\
&=&   (\sqrho^{n+1})^2 u_\alpha^{n+1} - (\sqrho^{n})^2 u_\alpha^{n}. 
\eeqn 
All other definitions of the time steps $n+a$, $n+b$, \dots $n+f$  can be chosen freely. By choosing $a,b, c, d, e= 0 $ a linear
 implicit scheme  can  be obtained. Using  $a,b, c, d, e= 1/2 $ with the time average of the 
pressure as $ p^{n+1/2}= ( p^{n+1} - p^{n})/2 $   gives
 the known scheme cited above. We stick to this non-linear implicit  scheme in this paper, as it is of second order.  

From (\ref{sqrBinom}) it can be seen, that we are free to choose to simulate  in $\sqrho$ or in $\rho$. In the first case we 
have to frequently calculate the square of the values, while in the second case we have to frequently extract the square root.
 As the first operation is usually numerically  much cheaper we prefer to use $\sqrho$ as a variable.

Higher order one step methods can be similarly constructed. A dedicated paper comparing different time integrators  
for skew-symmetric schemes is submitted \cite{Brouwer2013}. 

\section{Boundaries}
\label{boundaries} 
\label{secBound}

Boundaries have a special role in skew-symmetric schemes, as the properties so far essential in our discussion are broken.  
Both, the skew-symmetry and the telescoping sum property do no longer hold, if one-sided derivatives are used. 
The alternative approach of using ghost points will not be pursued. 
Instead, we exploit, that we can choose derivatives matrices largely arbitrarily,  by picking summation by parts (SBP) derivatives  \cite{Olsson1993,Strand1994}. 
SBP derivatives fit perfectly into a conservative FD scheme, as they lead to well defined fluxes at the boundaries.   

To introduce the concept of SBP derivatives consider the most simple boundary condition, 
 a slip wall.
 Here the  perpendicular velocity  vanish, implying  a vanishing mass flux; 
in one dimension the semi-discrete mass equation  is 
\beqn
    \partial_t  \rho +       D_x u   \rho  &=&0 .
\eeqn
Numerical fluxes at the boundary are introduced by the modified stencil at the boundary: they are again calculated 
by summing, using the same short hand notation as above 
\beqn
    \partial_t    \hone^T \rho +    \underbrace{\hone^T   D_x}_{b^T} u   \rho  &=&0.
\eeqn 
Choosing  the  most simple central second order derivative, which reduces to first order at the boundary, 
we obtain 
\beqn
b^T &=& \hone^T  D_x 
= 
\hone^T \frac 1 {\Delta x} \left(
\begin{array}{cccccc} 
-1 & 1 & \\
-\halb & 0 & \halb & \\
&-\halb & 0 & \halb & \\
&& \ddots & \ddots & \ddots   \\
&&& -1 & 1   
\end{array}\right)\no\\
&=&\frac 1 {\Delta x} ( -\frac 3 2 , \halb , 0 ,\cdots , 0, -\halb , \frac 3 2 )  ,
\eeqn
thus, the flux over the boundaries is 
\beqn
{b^T} u   \rho& = &  - \left( (\rho u )_1 \frac 3 2 - (\rho u )_2 \halb   \right)
   +    \left( (\rho u )_{N} \frac 3 2 - (\rho u )_{N-1} \halb   \right). 
\eeqn   
This is a consistent flux, as it reduces for constant values to $\pm (\rho u )$ on either side. 
However, a zero  velocity at the boundary $ ( u )_1= 0 $, does not imply a zero flux, but a flux of  $- (\rho u )_2 $,
 which is  a numerical artefact. A similar term arises for the internal energy, which possibly destabilizes the simulation.
 This could be eliminated by using ghost points or by prescribing the numerical flues instead of the boundary values. 

Another way to cure this problem is by changing the derivative, so that $ ( u )_1= 0 $ strictly 
implies zero flux. This demands, that  $b^T  = \hone^T  D  $ is zero for all 
elements but the first and the last, which are $\mp 1$ for consistency. Using further the consistency condition for 
a derivative $ D \hone = 0 $ we arrive at 
\beqn
W u' =   \left(
\begin{array}{cccccc} 
-\halb & \halb & \\
-\halb & 0 & \halb & \\
&-\halb & 0 & \halb & \\
&& \ddots & \ddots & \ddots &  \\
&&& -\halb & \halb   
\end{array}\right) u .\label{sbpM}
\eeqn
The matrix $W$ contains weighting factors on the diagonal, 
and is needed as the first and last line of $D$  represents only half of a derivative. Thus, the weights have to be 
\beqn
W_{ij} =  =\left\{ 
\begin{tabular}{cl} 
$\halb$& \mbox{\quad for \quad } $ i=j= 1,N   $ \\
1 &\mbox{\quad for \quad } $ i = j \not =1,N $ \\
0 &\mbox {\quad for \quad } $ i \not= j$ \\
\end{tabular}
\right. . 
\eeqn
This is the most simple example of a SBP derivative.    
Our scheme assumes explicit derivative matrices. Matrices with a diagonal norm,  as the example above, are easily included in the given scheme. 
Such SBP matrices of higher order are derived in the work of Strand \cite{Strand1994}. 
He derives such quasi-explicit derivatives, for which the order of the  discretization error is  half of the order in the interior.  
 The norm, or weights, can be simply absorbed into the Jacobian $\tilde J_{i_1,i_2} = \tilde J_{i_1,i_2} W^{\xi}_{i_1}W^{\eta}_{i_2} $, as the $\xi$-weight-matrices interchange with the $\eta$-derivative and vice versa. 
The multi-index $\bf i = (i_1,i_2) $ is for the $\xi$ and $\eta$ direction in calculation space.  

The flux is now simply calculated as before, and is in line with the analytical flux. For example the conservation of 
mass is 
\beqn
\sum_{\bf i} \tilde J_{\bf i} \rho^{n+1}_{\bf i} 
&-&  
\sum_{\bf i} \tilde J_{\bf i}  \rho^{n}_{\bf i}   \no \\
&&+    
  \Delta t 
\left(\frac {1}{\Delta \xi  } \sum_{i_2,i_3}  W^{\eta}_{ i_2} W^{\zeta}_{ i_3}  \tilde u_{\bf i} \rho_{\bf i}|_{i_1=1}^{i_1=N}  
     +\frac {1}{\Delta \eta } \sum_{i_1,i_3}  W^{\xi}_{ i_1}  W^{\zeta}_{ i_3} \tilde v_{\bf i} \rho_{\bf i}|_{i_2=1}^{i_2=N}  
     +\frac {1}{\Delta \zeta} \sum_{i_1,i_2}  W^{\xi}_{ i_1}  W^{\eta}_{ i_2} \tilde w_{\bf i} \rho_{\bf i}|_{i_3=1}^{i_3=N} 
 \right)
= f^{\rho}
\eeqn
The $\tilde u_\alpha$ are defined as in (\ref{uTilde}).   
The  flux  $ f^\rho$ appears if boundary conditions are enforced.  
Note that the expression, while a little bit clumsy,
 is very  close to the analytical expression. 
 Mass fluxes are given by the mass times velocity components  perpendicular to the boundary. 
 The $W^\alpha $ appear as integration weights in the volume and surface integrals.     

Also the momentum and energy flux at the boundary reduces in the same manner to the analytical flux;  here, beside
 the special form of the flux  weights $b$ of the divergence terms, also skew-symmetry, with the exception of 
  the elements $D_{1,1}= - D_{N,N}= \halb$, is needed; this leads to the correct fluxes of the kinetic energy over the boundary.   

Enforcing the boundary conditions can be done in different ways. We use in the following a characteristic splitting 
and set the boundary values accordingly by overwriting. 

An interesting alternative  to set boundary conditions is presented by Carpenter et al. \cite{Carpenter1994220}, which allows to derive strict energy (norm) estimates. 
However, this approach is formulated in the  characteristic form of the equations and needs to be adopted for the form used here.   

\section{Numerical examples}
\label{numericalExamples}
 
In the following we present two numerical examples which span the intended area of application: shocks and acoustics. The simulations are done in  matlab  and are thus limited in size. Large parallel computations in a Fortran implementation are ongoing, which will cover the interplay between these two aspects. First results are presented in  \cite{Brouwer2013}.  

\subsection{Non-linear Pressure Pulse} 

\begin{figure}
\includegraphics[  width=0.46\linewidth ]{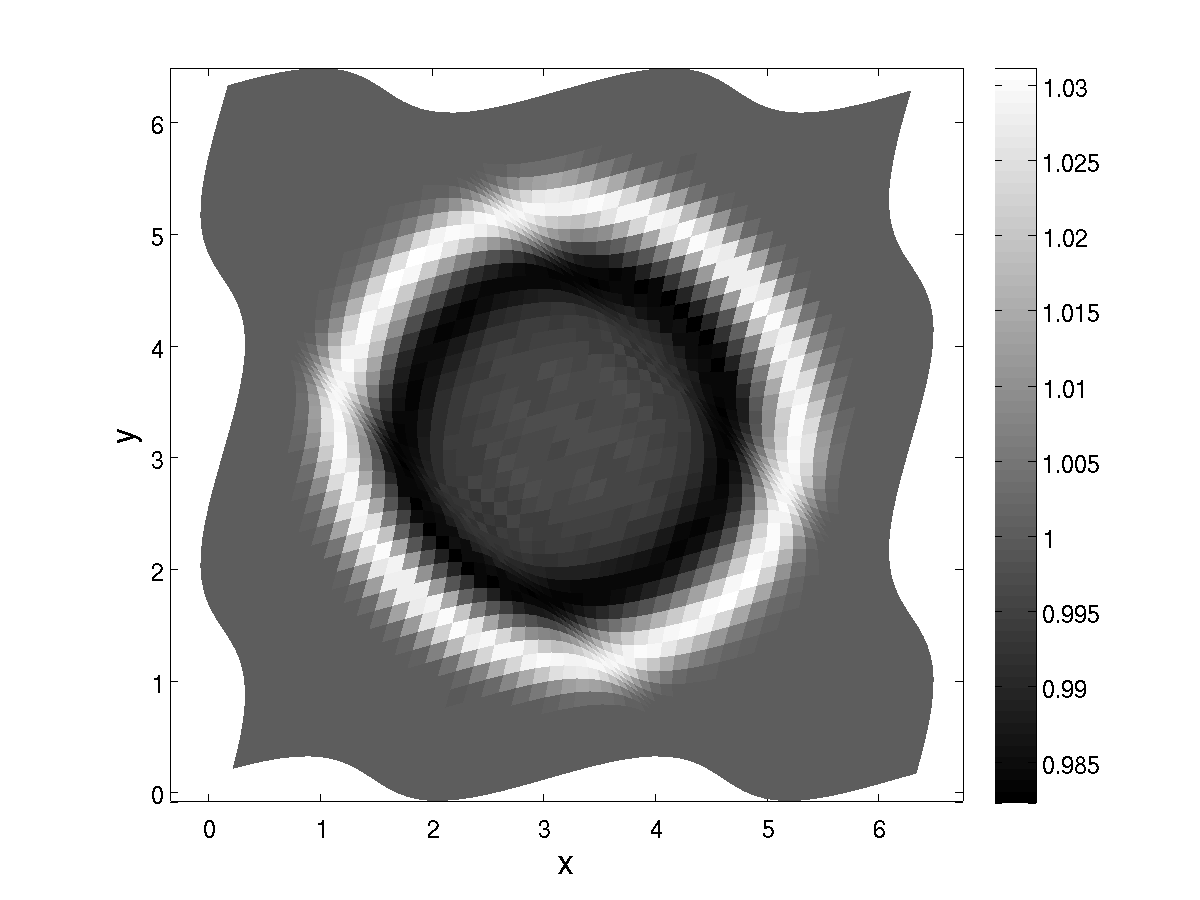}
\includegraphics[  width=0.46\linewidth ]{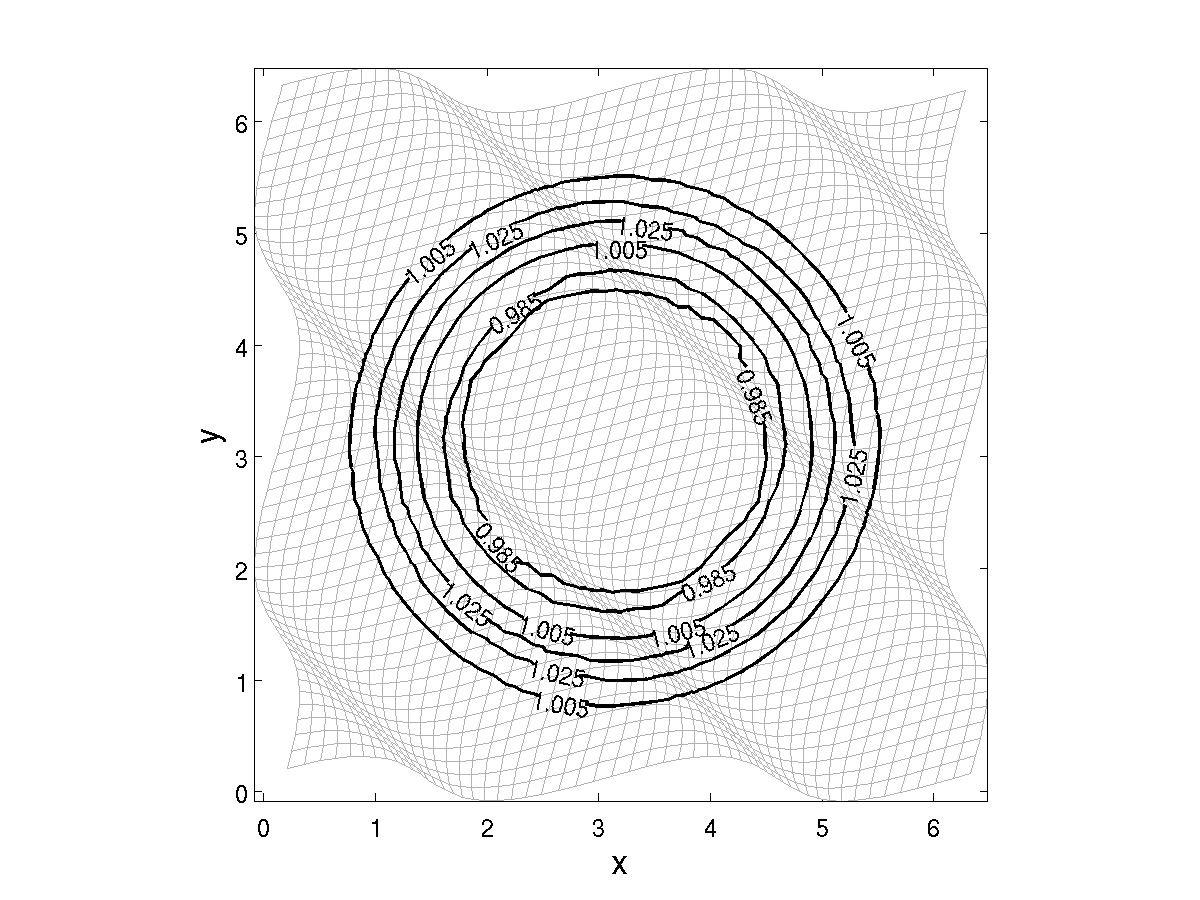}
\caption{The density at T = 0.005s of an adiabatic pulse on a strongly distorted grid. Not that the correct interpolation of contour lines might add difficulties.}
\label{pulseGray}
\end{figure}

\begin{figure}
\includegraphics[  width=0.46\linewidth ]{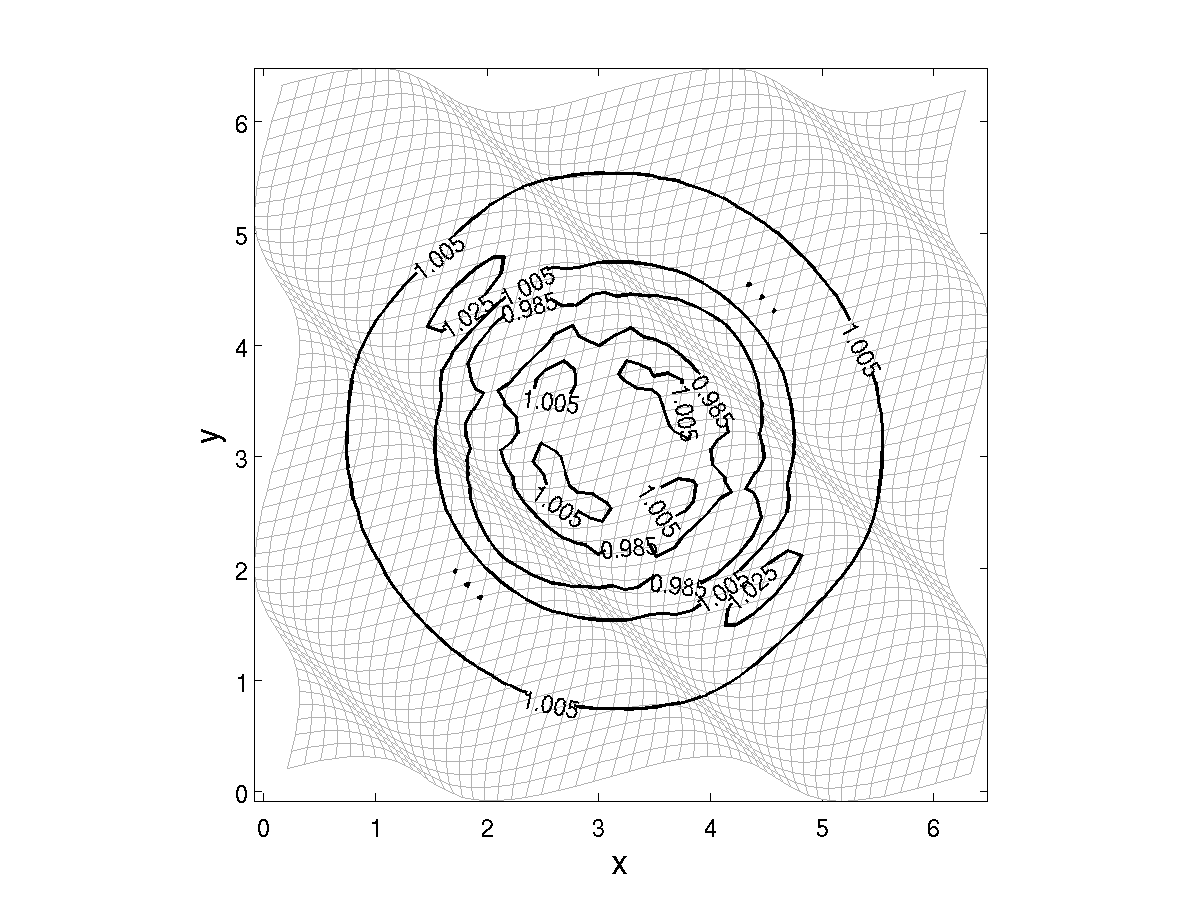}
\includegraphics[  width=0.46\linewidth ]{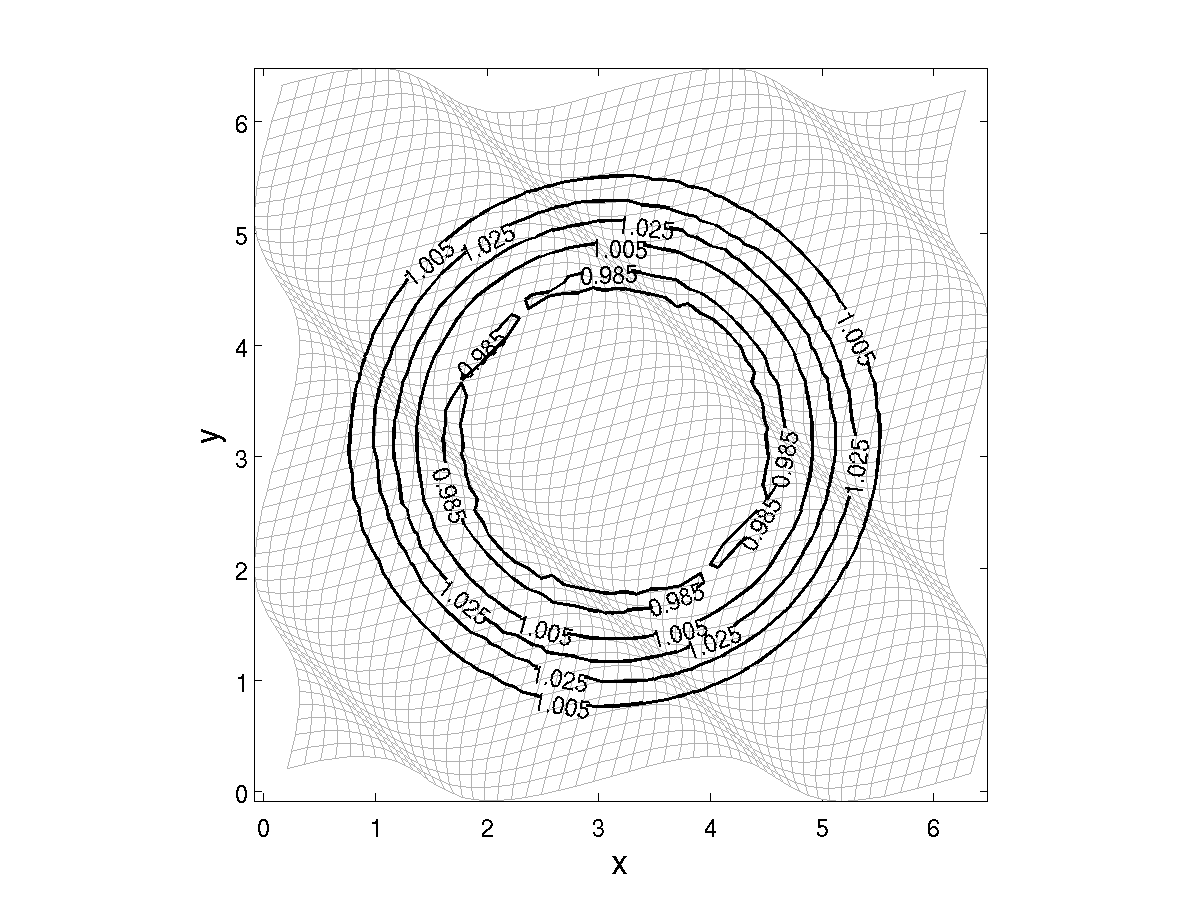}
\caption{Selected contour lines of the density at T = 0.005s for a second order derivative and the Tam and Webb 4$^{th}$ order derivative. The second order derivative has strong artefacts, while the TW derivative is much better.}
\label{pulseCont}
\end{figure}

\begin{figure}
\includegraphics[  width=0.46\linewidth ]{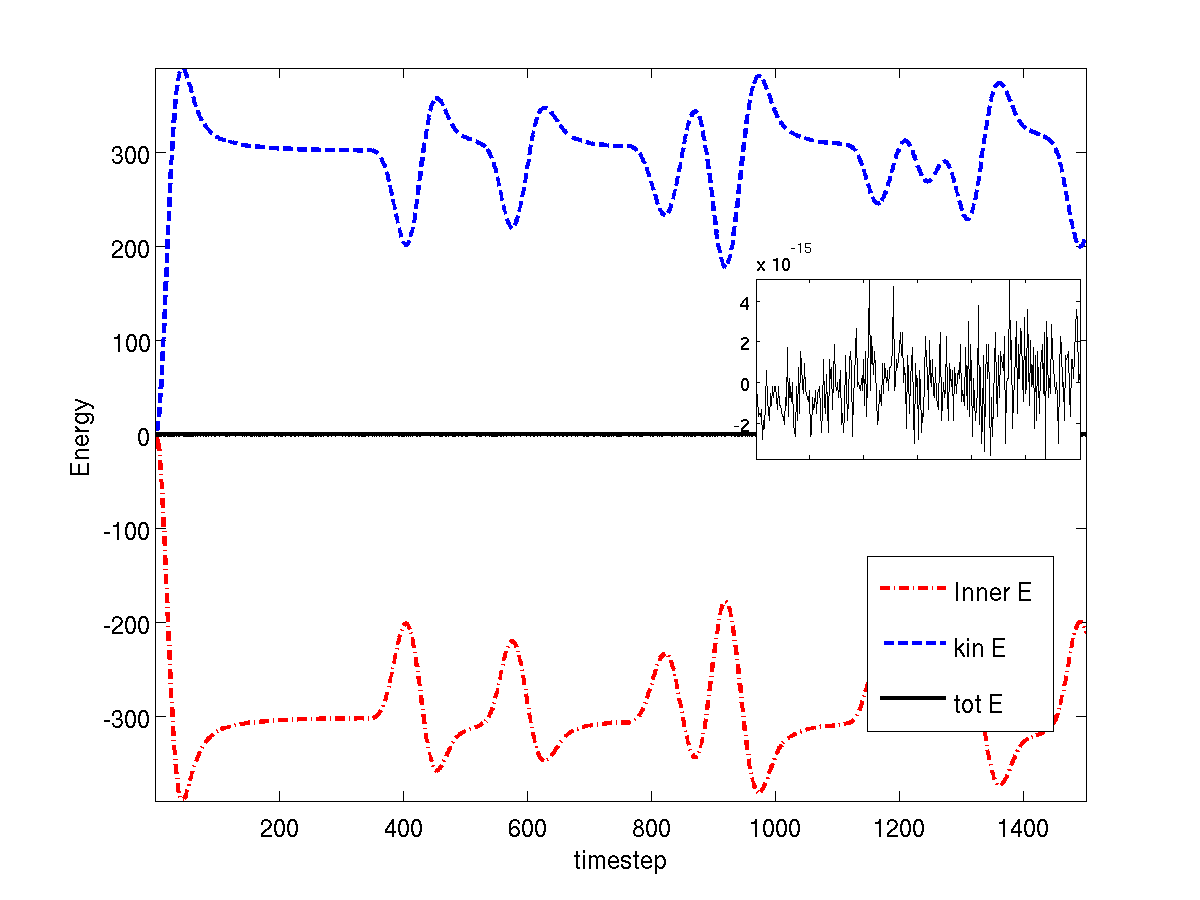}
\caption{The conservation of the energy, calculated from mass, velocity and pressure. The change is at machine precision.}
\label{energyPeriocic}
\end{figure}

The figures (\ref{pulseGray}-\ref{energyPeriocic}) show as an example an adiabatic pressure pulse, with $\rho_{peak} = 1.25 \rho_0$, $\rho_0=1kg/m^3$, $p_0=10^{5}Pa$, $u_0=v_0=0$. The grid is periodic with $x=\xi+ 0.2\sin(k(\xi+\eta))$, $y=\eta+ 0.2\sin(k(\xi+\eta))$, $k=2$. The values are chosen to create a very bad, i.e. extremely distorted 
grid for test purposes.
The time step is  $\Delta t =2\cdot 10^{-5}s$, at a resolution of $55\times54$ grid points,  and the boundaries are periodic. The time stepping is done by the adapted implicit midpoint rule. We find that for high order derivatives the distortion is having little effect on the result quality see fig. (\ref{pulseGray}). In comparison the performance of 
second order derivative is much worse, while the Tamm and Webb derivative creates reasonable results.   
Despite being in the  the non-linear regime, we find energy, mass and momentum conservation up to machine precision, cf. 
 fig. (\ref{energyPeriocic}). We want to stress that we do the calculation in primitive variables from which the energy
 is created by 
$
   \sum_{\bf i} J_{\bf i } 
   \left( 
      \frac {p _{\bf i }}{1-\gamma} 
       + 
      \rho_{\bf i}\frac { u_{\bf i }^2 +u_{\bf i }^2 } {2}
    \right)\Delta \xi \Delta \eta.
$   

\begin{figure}
\includegraphics[  width=0.46\linewidth ]{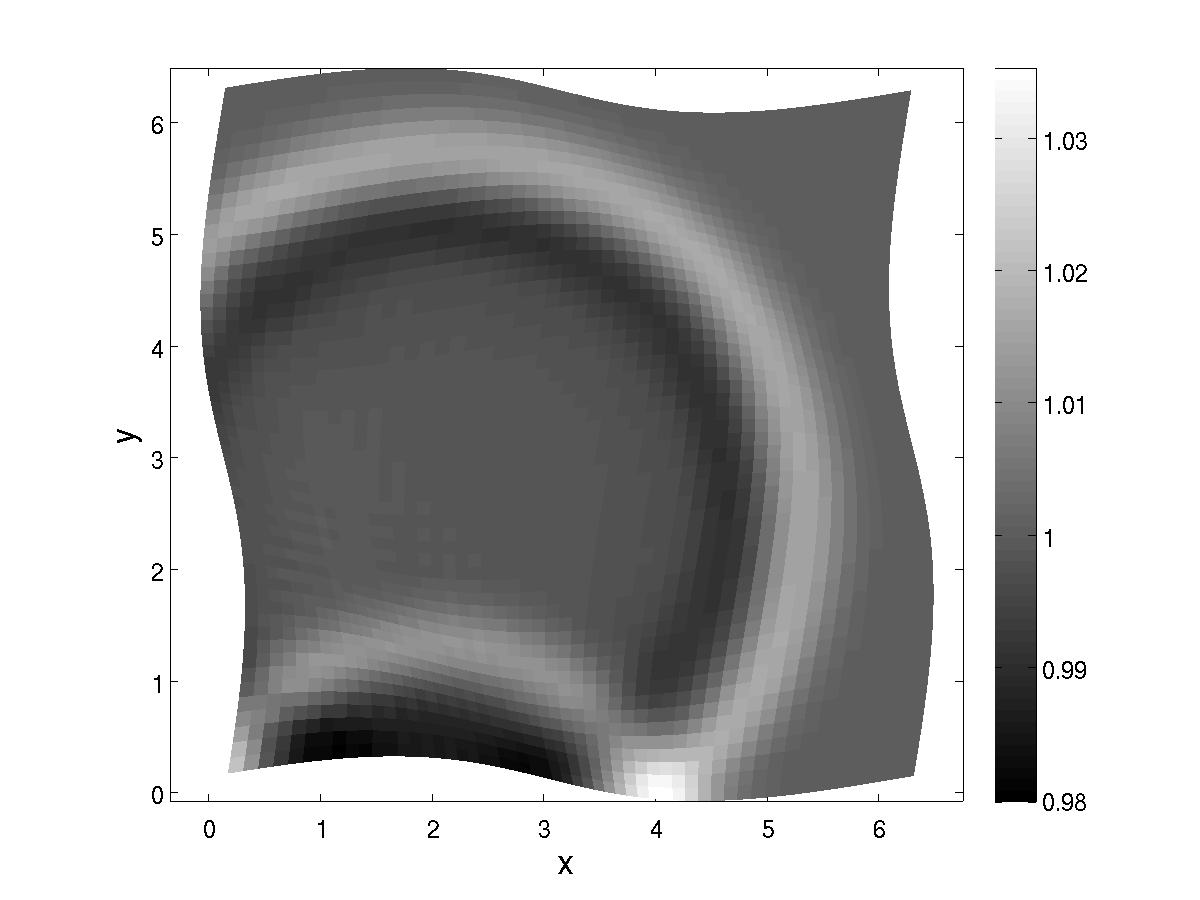}
\includegraphics[  width=0.46\linewidth ]{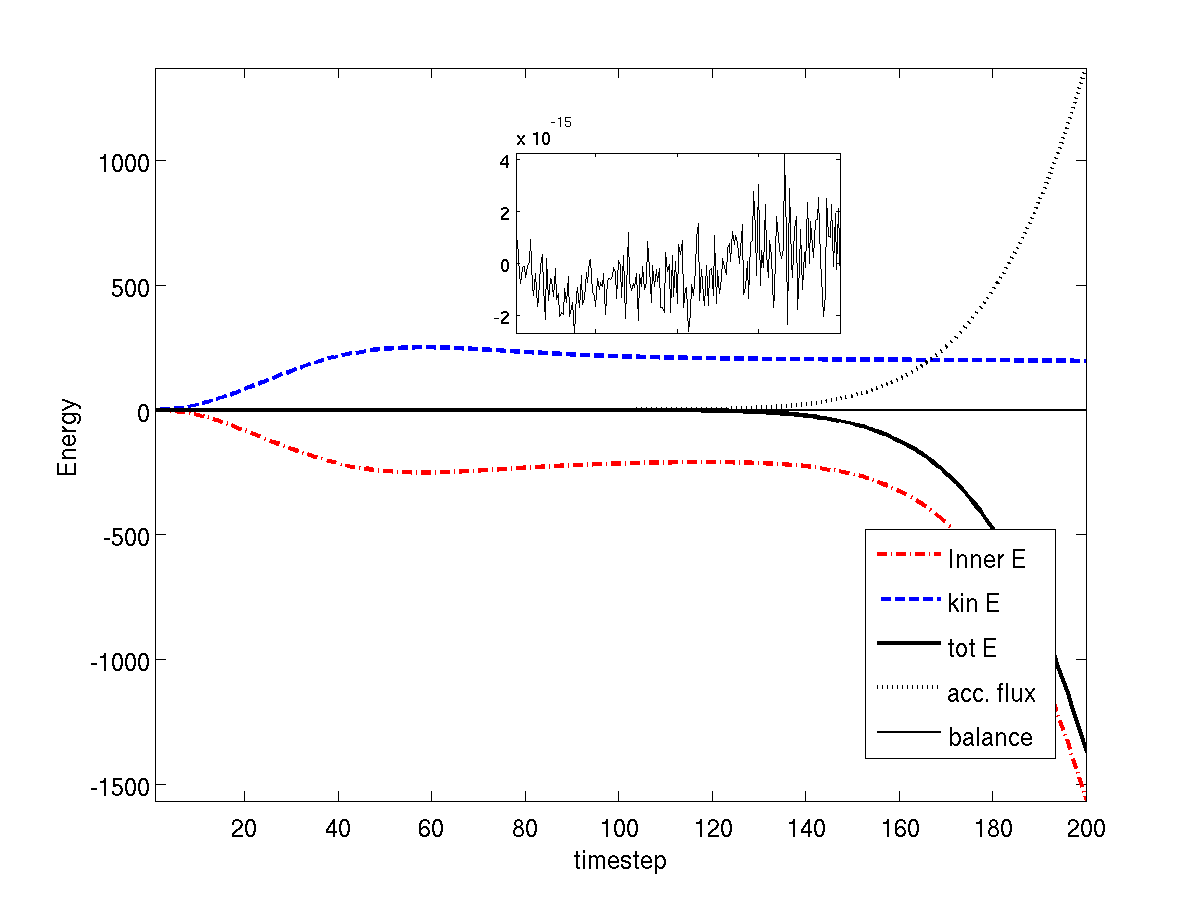}
\caption{The density an adiabatic pulse on a non-periodic grid.  
Left: Accounting for the energy flux over the boundaries the energy is again conserved up to machine precision.}
\label{pulseGrayOpen}
\end{figure}

The same pressure pulse for a  non-periodic grid with $k=1$ is presented in fig. \ref{pulseGrayOpen}; the boundaries are a slip 
wall and an non-reflecting boundary. 

 The boundaries conditions are calculated by a split in characteristic variables. 
 The derivative is standard fourth order in the interior and reduces to a second order derivative described by Strand \cite{Strand1994}.
  Fluxes are  therefore calculated from boundary points only;  enforcement of  boundary conditions leads to a small extra term, 
     which can uniquely be calculated by the difference
 between the calculated next time step without boundary conditions, compared with the values after setting the boundary conditions.  
 Including the accumulated  fluxes leads to a total conservation down to machine precision as in the periodic case.        

\subsection{Shock configuration}

In the following  we present a test case given by Lax and Liu \cite{LaxLiu1998}. Test case 13 was chosen, as shock and slip lines are present. 
We use no physical damping. 
As the scheme does not have numerical damping, we have to include a dissipation mechanism. 
This allows us to carefully add only the dissipation, which is needed on physical grounds, as in this example, demanded by shocks. 
A good way to do this is the {\it adaptive} and conservative  filtering, presented by Bogey et al. \cite{Bogey2009}, 
which depends on the divergence of the flow field. By this we introduce dissipation only where it is needed. 
 This filtering is conservative when applied to conserved quantities, thus we filter the mass $\rho$, the components of momentum $\rho u_\alpha$ and the internal energy $p/(\gamma-1)$,  which we increase by the change of kinetic energy to mimic a physical dissipation mechanism. 
 By this we preserve the perfect conservation as before. 
The filtering is done after every time step. 
The adaptive filter can in principle create noise by changing the filter strength from time step to another. 
Even though this was not an issue in this test case, which proved to be very robust within our approach, 
we use a soft switching function for the filter strength $\sigma$ determined by the output of a shock detector $r$
\beqn
\sigma = 1-\mbox{tanh}((r_{th}/r )/\lambda ).
\eeqn
instead of the more abrupt function used in \cite{Bogey2009}. 
A full description of the filter can be found in the given reference. 
The steepness of the switching function was $\lambda= 2$ in the simulation, the threshold was $r_{th}=10^{-5}$.
 The initial values for $(\rho,u,v,p)$ are   (upper-right, upper-left, lower-left, lower-right) 
(1, 0,  -0.3, 1), (2, 0, 0.3, 1),  (1.0625, 0, 0.8145, 0.4) and (0.5313, 0, 0.4276, 0.4). 
As boundary conditions standard, non-reflecting  boundary conditions are used, where the reference values are 
taken as the values in the neighbouring grid-points in the interior of the domain. 
\begin{figure}
\includegraphics[  width=0.46\linewidth ]{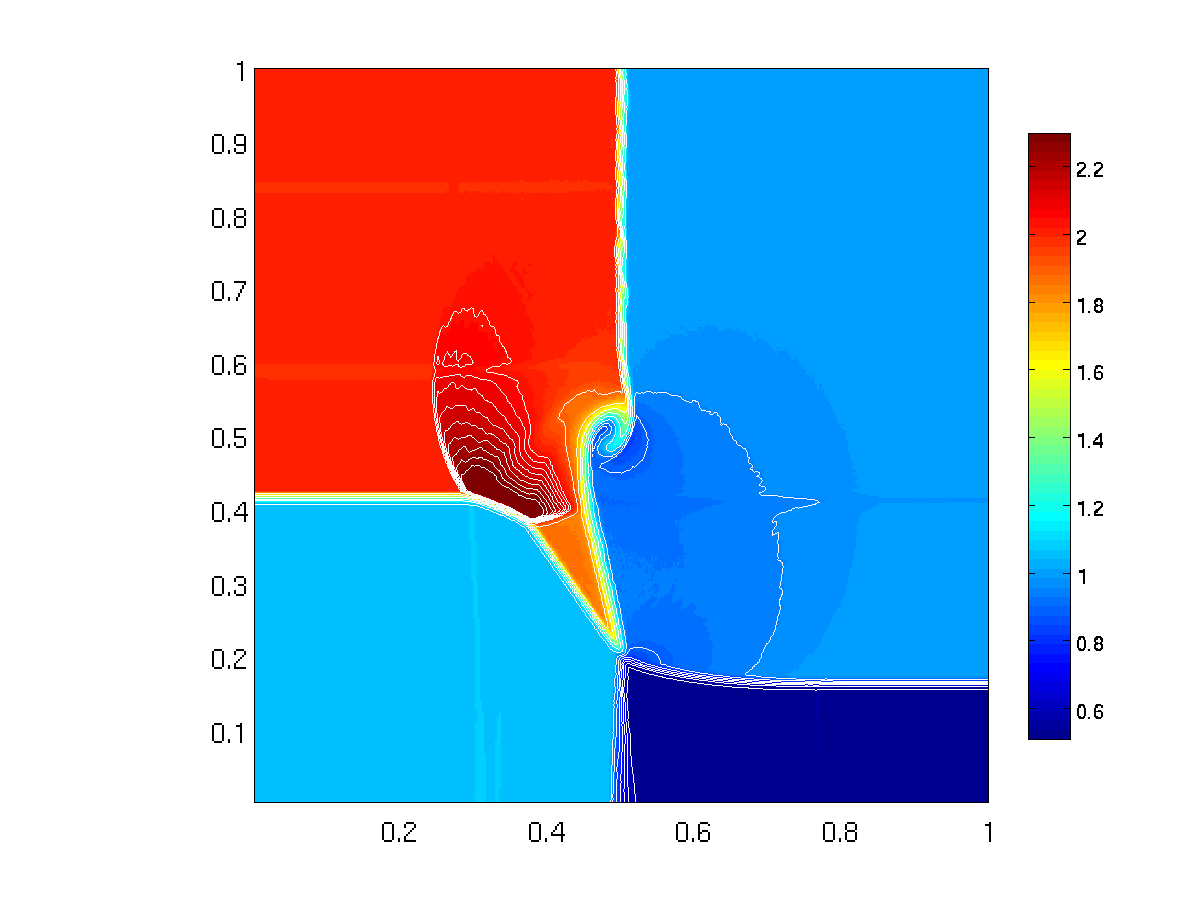}
\caption{The density at t = 0.3 of the shock test case 13 of  Lax and Liu \cite{LaxLiu1998}. Overall very good agreement is found, see discussion in the text for details.}
\label{LaxLiu13}
\end{figure}

\begin{figure}
\begin{center}
\includegraphics[ clip=true, width=0.505\linewidth ]{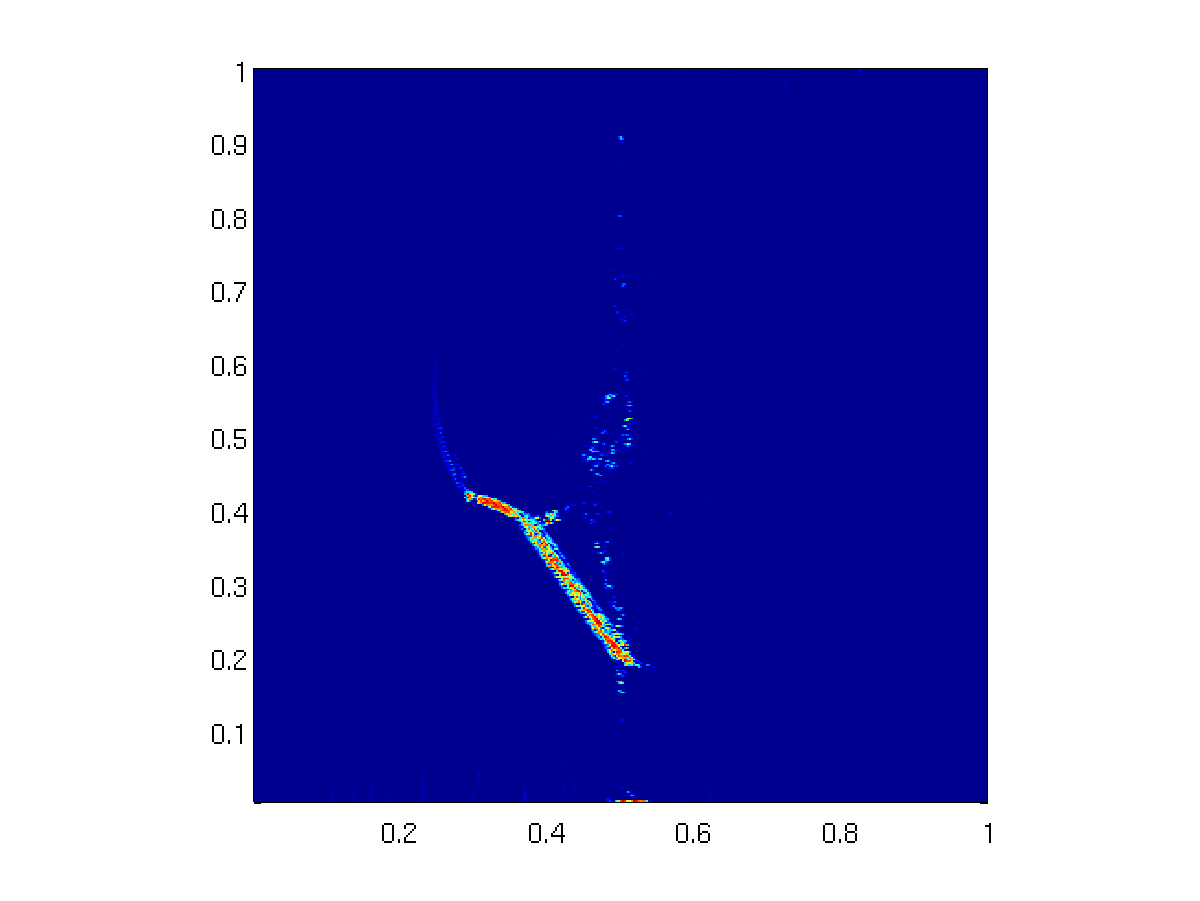}
\includegraphics[ clip=true, width=0.485\linewidth ]{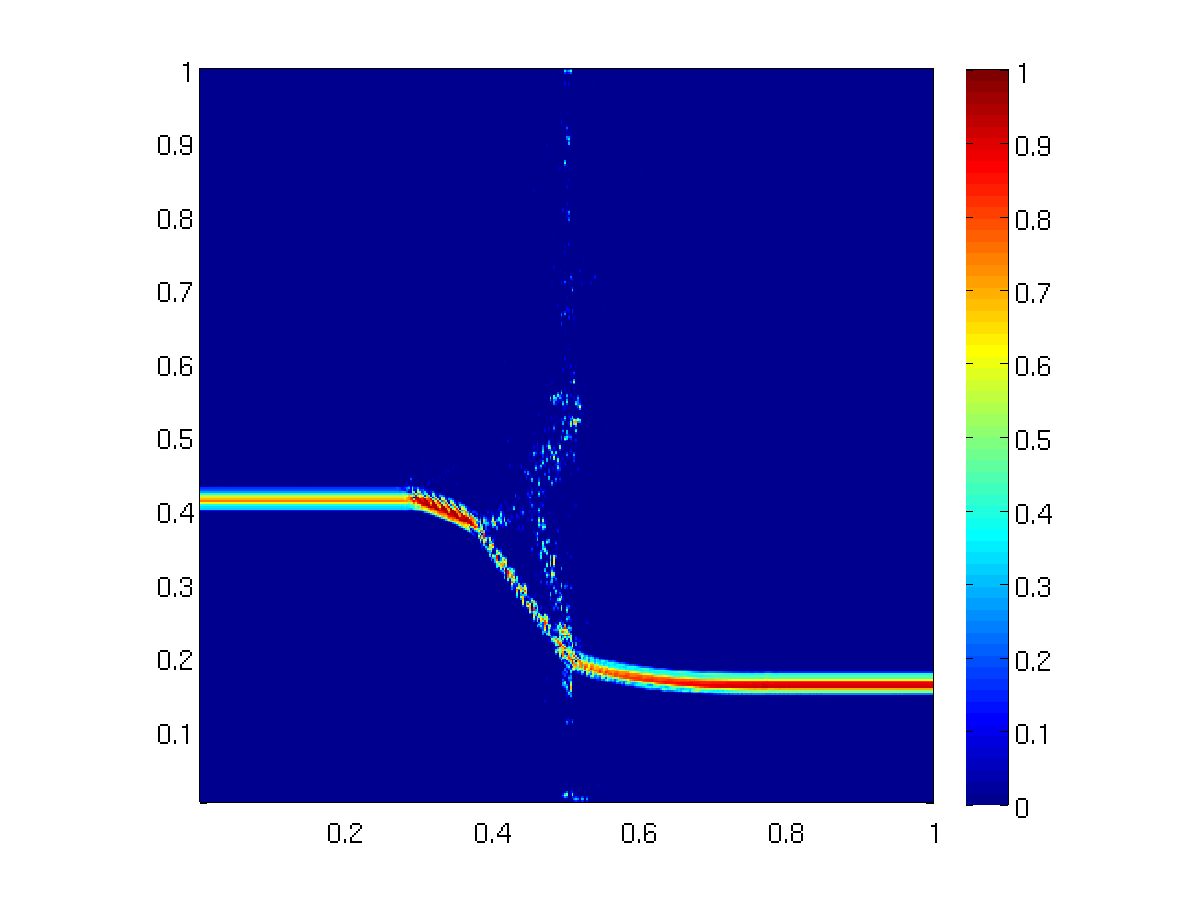}
\end{center}
\caption{The adaptive shock filter strength $\sigma$ in x and y direction. It is apparent, that only at shocks the filter is involved whereas in other regions is is nearly zero.}
\label{ShockFilter}
\end{figure}

Figure (\ref{LaxLiu13}) shows excellent agreement of the density with the reference solution. 
The shock speeds agree with the reference case, as with the analytical formula; this is expected for a locally conservative code. 
 For example the shock between upper-right and lower-right the shock speed is calculated as $s = [\rho v^2+p]/[\rho v] $ to be $s= -1.1246$, giving a shock location of $y= 0.1626$. This is in perfect agreement with the simulation.
To guarantee correct shock speeds, the scheme should have local and consistent fluxes. This issue is discussed for our FD scheme in \ref{AppFluxes}.  
However, three differences are visible. 
 First the contour levels are not as smooth as in the original work; this is attributed to the fact that away from shocks absolutely no dissipation is involved. 
 This could easily be smoothed by adding a small amount of physical or numerical dissipation, but it was left frictionless on purpose in this test case. 
 Further the roll up in the center seems slightly different, and the slip line in the upper half is weakly wrinkled. 
 The roll up was observed to depend sensitively on the chosen dissipation and might thus be hard to reproduce, as the {\it implied} friction in code used in \cite{LaxLiu1998} is difficult to estimate. 
 The wrinkled slip line can be identified as the onset of a Kelvin-Helmholtz instability. 
 This might be suppressed in \cite {LaxLiu1998} by the implied numerical damping. 
In contrast to the reference case, the dissipation in this simulation is the adaptive filter strength and is therefore known. 
It is shown in fig. (\ref{ShockFilter}), which shows that, whereas at  shocks strong dissipation is found, the slip lines are  only weakly filtered.

\section{Conclusion} 

We presented a skew-symmetric finite difference scheme for fully compressible flows. 
It is strictly skew-symmetric and conservative also on distorted grids. 
The discrete form does purely build from point-wise operations and derivatives. 
This allows to use any explicit derivative which is skew-symmetric and has the telescoping sum property, in effect all derivatives with a symmetric stencil. 
By this any discretization order and wave-number optimized derivatives can directly be used. The implementation is simple and numerically efficient. 
Due to the described properties acoustics and shocks can be treated equally well. Boundary conditions and fluxes at the boundary can neatly be treated
by summation by parts derivatives.      

Appropriate time stepping schemes of high order and the use of standard time integrators 
are presented in a separate paper \cite{Brouwer2013}.  


\appendix

\section{Comparison with skew-symmetric schemes utilizing averaging}
 \label{appendixCompare}
 	 
It is interesting to investigate, how this scheme compares with other skew-symmetric schemes, especially to the classical ansatz by Morinishi et al. 
\cite{Morinishi1998}, starting point for several other works on compressible and incompressible flows. 
Their discrete skew-symmetric momentum transport
 term, for second order accuracy, is in  their notation, where we use Greek letters for the space index   
 $
2\cdot(Skew) = (Div)+(Adv)=
\frac {\delta_1}{\delta_1 x_\beta} \left({\overline u}^{1x_\beta}_\beta {\overline u}^{1x_\beta}_\alpha\right)
+ \overline {{\overline u}^{1x_\beta}_\beta \frac {\delta_1}{\delta_1 x_\beta} {\overline u}^{1x_\beta}_\alpha}^{1x_\beta}
$.
The averaging operation is
$
{\overline \phi}^{nx_\beta}|_{\bf x } =\left(
\phi |_{{\bf x} + \frac n2 {\bf e}_\beta  }
+\phi |_{{\bf x} - \frac n2 {\bf e}_\beta  }
\right) /2
$
and the derivative is
$
\frac {\delta_n}{\delta_n x_\beta}\phi
=\left(
\phi |_{{\bf x} + \frac n2 {\bf e}_\beta  }
-\phi |_{{\bf x} - \frac n2 {\bf e}_\beta  }
\right)/(nh) 
$
with the unit grid vector ${\bf e}_\beta$ and the grid spacing $h$. 
A straight forward calculation reduces to the expression used by us (with $\rho\equiv 1$): 
\beqn
2(Skew)& =& \frac{
u_{\alpha, {\bf x}+ {\bf e}_\beta} u_{\beta, {\bf x}+ {\bf e}_\beta} 
-
u_{\alpha, {\bf x}- {\bf e}_\beta} u_{\beta, {\bf x}- {\bf e}_\beta} 
}{2h} 
+
 u_{\beta, {\bf x}+ {\bf e}_\beta} 
\frac{
u_{\alpha, {\bf x}+ {\bf e}_\beta}
-u_{\alpha, {\bf x}- {\bf e}_\beta}
}{2h}\no\\
&=& D_{x_\beta} (u_\beta u_\alpha) -u_{\beta}D_{x_\beta}u_\alpha  =(D_{x_\beta}U_\beta \cdot + U_\beta D_{x_\beta} )u_\alpha
\eeqn
Choosing $\beta = 1 $ and $u_\alpha = u$ for the one dimensional case we arrive at our expression above. 
One should note, that the advective and the divergence term {\it are different to ours}  due to the special 	 averaging,  
and that {\it only the combination} are the same due to some cancellation and addition. 
A similar behaviour is expected for higher order derivatives. 
A generalization of this averaging procedure to the compressible case is presented by Kok \cite{Kok2009}. The full scheme is expected to be different,
 as the other terms like mass flux are also constructed by averaging and no combination seems to reduce those to our form.    

Our  understanding is, that the spirit behind these works \cite{Morinishi1998,Kok2009} is  to  use averaging procedures 
to obtain a discrete version of the product rule, while we avoid to depend on this rule in the discrete by rewriting  the equations analytically in a proper form.
%

\section{Defining Fluxes}
\label{AppFluxes}

When simulating  shocks wrong shock speeds can be found even for consistent schemes \cite{LeVeque1992}. 
The  Lax-Wendroff theorem provides a set of conditions for which proper shocks are guaranteed. 
An essential condition for this theorem are  
local and consistent fluxes; a numerical flux is said to be local if it is calculated from a {\it finite} number of neighbouring discretization values;    
a numerical flux is consistent if it reduces to the analytical flux when this neighbouring discretization values are chosen to be constant.    
Being able to define consistent and local fluxes is therefore of central importance. 

However, we stress , that the fluxes are not explicitly used in the implementation; it is a standard FD scheme building  purely 
on derivative operations for means of simplicity and efficiency. The following part is thus only of theoretical importance.

We derive fluxes in one dimension for the sake of brevity; the derivation can be easily extended to more dimensions.
We also work in time continuous case, as our goal are the spatial fluxes. 
 A simple way to derive the fluxes is to split the 
sum over the full space occurring in the derivation of the global conservation into two parts, thus the derivation of the 
fluxes follows  the proof 
of the conservations closely. This splitting written 
in the vector notation introduced earlier is $\hone = \hl_{j+\frac 1 2} + \hr_{j+\frac 1 2} $, where the $i^{th}$ component of vector $\hr_{j+\frac 1 2}$  is given by
$$
\hr_{j+\frac 1 2} =\left\{ 
\begin{tabular}{ccl} 
$0$ &\mbox{ for  }& $ i \leq j$ \\
1   &\mbox{ for  }& $ i \geq j+1$ 
\end{tabular}
\right. , 
$$  
and $\hl_{j+1/2}$ accordingly with interchanged zeros and ones.  The mass flux is derived from 
\beqn
&&- \hone^T 2\sqrho \partial_t  \sqrho  = 
 - \partial_t \hone^T    \sqrho^2  \no\\
 &=& (\hone)^T  D (\rho u )  =   (\hl_{j+\frac 1 2}+\hr_{j+\frac 1 2})^T  D (\rho u )  .
\eeqn
 We are interested in the flux in the interior of the domain, we can thus assume zero fluxes at the domain-boundary and therefore conservation of mass. The  telescoping 
sum property of the derivative therefore leads to 
\beqn
0  =  \underbrace{\hl^T_{j+\frac 1 2}  D (\rho u )}_{f_{j+1/2}} + \underbrace{\hr^T_{j+\frac 1 2}  D (\rho u )}_{f_{(j+1)-1/2} = -f_{j+1/2}}.
\eeqn
We can already conclude, first, that we can define fluxes and that secondly we have local fluxes, since we assume a finite stencil with constant 
coefficients\footnote{This can be easily chosen, as grid transformations are used for non-equidistant grids. The direct use of compact derivatives could 
invalidate this assumptions.}. 
To prove the consistency  of the fluxes  we need  information about the coefficients of the derivatives.
 Thus we calculate the weight coefficients $ b_{j+\frac 1 2} $ of the the flux, given by  
\beqn
&&\hr^T_{j+\frac 1 2} D (\rho u ) =  b_{j+\frac 1 2}^T  (\rho u ).
\eeqn
The zeros in the definition of $\hl_{j+\frac 1 2}$  remove the upper  part $m \leq j$ of $D_{m,n}$, thus
$ b_{j+\frac 1 2} $ is given by 
\beqn    
 b_{j+\frac 1 2} &=&\hone^T\left(
\begin{tabular}{cccc|cccccc} 
  \multicolumn{3}{c}{0} && \multicolumn{5}{c}{0} \\
  \hline
& $-a_{3}$ &$ -a_{2} $ &$-a_{1} $ & 0 &$ a_1$& $a_2$ &$a_3$  \\
&$\ddots$& $-a_{3}$ &$  -a_{2}$ &$-a_{1} $ &$ 0 $ &$ a_1$ &$ a_2$ &$a_3$   \\
&&$\ddots$& $-a_{3}$ &$  -a_{2}$ &$-a_{1} $ &$ 0 $ &$ a_1$ &$ a_2$&$\ddots$   \\
& &&&$-a_{3}$  &  $\ddots$ &$\ddots$ \\ 
\end{tabular}\right)\no\\
& =& ~~- ( 0,A_n, \dots, A_3,A_2,A_1,A_1,A_2,A_3, \dots) \label{sumrD} .
\eeqn
with $A_k=\sum_{i=k}^n a_{i}$. The finite stencil width is $n$. Note the symmetric structure of this coefficients.  
Therefore the mass flux is 
\beqn
f_{j+\halb}^{\rho} = b_{j+\halb}^T (\rho u)=\sum_{k=1}^n A_k 
\left( (\rho u)_{j+k} + (\rho u)_{j-k+1}  \right)  \label{fluxRho} \\
= \sum_{i=1}^n a_i \sum_{m=0}^{i-1} \left( (\rho u)_{j-m}+(\rho u)_{j-m +i } \right)  . 
\eeqn
The second form is gained by resorting the terms, also seen directly from the form of the matrix (\ref{sumrD}) .
If we assume $(\rho u)_{j+k}=const$ we get by summing of (\ref{fluxRho}) simply by inspection of  (\ref{sumrD}), 
\beqn
f_{j+\halb}^{\rho} = b_{j+\halb}^T (\rho u) =(\rho u) \sum_{k=1}^n A_k  = \rho u \underbrace{2 \sum\nolimits_{l} l a_l}_{=1}  = \rho u.
\eeqn   
The consistency condition of any numerical derivative enforces that the sum is unity.  

The momentum flux is calculated from the combination of the mass and the momentum equation
($\hone^T [  \ref{moriMom} +\halb U (\ref{moriMass}) ] $) 
$$
 \hone^T   \partial_t (u\Sqrho)  \sqrt{\rho} \no\\
+     \hone^T 
D^{\rho u}
u
  + \hone^T D  p 
   \no\\ 
 + \halb \hone^T  U (     D U \rho   )   =0.
$$
 Most terms in the momentum equation
 are in divergence form and can reduce to fluxes as before. 
Splitting the sum  $\hone = \hl_{j+\frac 1 2} + \hr_{j+\frac 1 2} $ and assuming zero flux over the boundaries we obtain
\beqn
%
 f^{\rho u}_{j+\halb}  = 
         b_{j+\frac 1 2}^T \left( \halb  ({ u}\rho)   u  + p 
\right)  
 + \halb \hr^T_{j+\frac 1 2} U [     D U \rho  + R    D u ]   .
\eeqn
The remaining term of the non-linear transport combines with the term in of the mass equation   
the last term is found to be 
\beqn
&& \hr^T_{j+\frac 1 2} U [   D U\rho  + R    D u ] =
 u^T \underbrace{\left[ ( \hr_{j+\frac 1 2}  D)^T+ ( \hr_{j+\frac 1 2}  D)  \right]}_{B} ({ u}\rho) \label{defB},
\eeqn
where the matrix $B_{m,n}$ has zero block for $m,n\leq j $ and $j+1 \geq m,n$ :
\beqn
B &=& 
\left(
\begin{tabular}{cccc|cccccc} 
&&&  &  $-a_{3}$ &$\ddots  $ &$  $ \\
&&& &  $-a_{2}$ &$ -a_{3} $ &$\ddots $  \\
&&& &  $-a_{1}$ &$ -a_{2} $ &$-a_{3} $ \\ 
  \hline
&  $-a_{3}$ &$ -a_{2} $ &$-a_{1} $ &  & &  &   \\
&$\ddots$& $-a_{3}$ &$  -a_{2}$ & & &    \\
&&$\ddots$& $-a_{3}$ &     
\end{tabular}\right),\no 
\eeqn
so that 
\beqn
u^T {B} ({ u}\rho) \no
&=& \sum_{i=1}^n  a_i  \sum_{m=0}^{i-1} \left( u_{j-m+i} (\rho u)_{j-m} + u_{j-m} (\rho u)_{j-m+i} \right)  .
\eeqn 
Again by inspection or summing it is easy to see that this numerical flux reduces to the analytical flux
$u (\rho u) $ provided we use a consistent derivative.

Altogether we obtain 
\beqn
 f^{\rho u}_{j+\halb}
 =
         b_{j+\frac 1 2}^T \left( \halb  ({ u}\rho u)     + p 
\right)   
 + \halb u^T {B} ({ u}\rho) ,\no  
\eeqn
where $\rho,u,p$ are the vectors of the discretization values. 
 
 The momentum flux by the non-linear transport  is therefore 
\beqn
&& u^T {B} ({ u}\rho) + b_{j+\frac 1 2}^T \halb  ({ u}\rho u)  \no\\
&=&\sum_{i=1}^n a_i \sum_{m=0}^{i-1} \left( (\rho uu)_{j-m}+(\rho uu)_{j-m +i } \right)  
+\sum_{i=1}^n  a_i  \sum_{m=0}^{i-1} \left( u_{j-m+i} (\rho u)_{j-m} + u_{j-m} (\rho u)_{j-m+i} \right) 
\no \\
&=&\sum_{i=1}^n  a_i  \sum_{m=0}^{i-1} \left[u_{j-m+i} + u_{j-m}  \right] \left[ (\rho u)_{j-m+i} +(\rho u)_{j-m}  \right] ,
\eeqn
which is just the same as in \cite{Pirozzoli2010}, which is {\it not} by chance. In the derivation of the momentum
 flux the momentum and the mass equation had to be 
combined, which leads to exactly the same spatial terms as in the flux splitting approach.  Of course 
differences in the two schemes arise when the energy equation is not split in the kinetic 
and internal energy parts, as it is apparently the case, and of course 
for the time discrete case, as a standard time integration scheme is used. 


The flux of the total energy is derived as $ \hone^T(\ref{moriEn})$ , where  $UD   p$ is replaced with the help of (\ref{moriMom})
\beqn
0&=& \partial_t  \hone^T p ~\phantom{(+ \frac{u^2}2)}+ \frac {\gamma}{\gamma-1}  \hone^T  DU  p -   \hone^T UD   p \\ 
 &=&\partial_t  \hone^T (p  + \frac{u^2}2)  + \frac {\gamma}{\gamma-1}  \hone^T  DU  p +  \hone^T U D^{\rho u }   u , 
\eeqn
and splitting the sum,  $\hone = \hl_{j+\frac 1 2} + \hr_{j+\frac 1 2} $, as before.  
The flux in the inner energy $ e = p/(\gamma -1)$ is analogous to the flux of the mass. The flux form the coupling to  the kinetic energy
$$
 \hr^T_{j+\frac 1 2} U  D^{\rho u } u    =  u^T   \tilde D^{\rho u } u     
$$
with $\tilde D^{\rho u } = \tilde DUR +RU\tilde D$, where $\tilde D_{m,n}$ is the derivative matrix, with the lines $m\leq j $ is set to zero. 
The  value of a  quadratic forms is given by the symmetric part of the matrix, here 
\beqn
\frac {(\tilde D^{\rho u })+(\tilde D^{\rho u })^T         }{2} = 
\halb \left( B      UR + RUB   \right)     
\eeqn
with $B$ as before. Thus the flux is 
$$
 f^{e}_{j+\halb}
 =
    \frac {\gamma}{\gamma-1}      b_{\hr,j+\frac 1 2}^T ( p u ) 
+ \halb u^T \left( B      UR + RUB   \right) u      
.  
$$
By the same arguments as before the flux is 
 local and consistent.

The derivation of fluxes for the three dimensional case on transformed grids are in principle calculated in the same manner, but the
 derivation is much  more cumbersome. 
The Jacobian is  multiplied to the conserved quantities, as expected. The geometrical factors for the conservative form are just multiplied.
 The geometrical factors  for the non-conservative forms enter in the quadratic forms and can be absorbed in the weight factors. 





\bibliographystyle{plain} 
\bibliography{../../../allRefsSince2008}{}

\begin{thebibliography}{10}

\bibitem{Bogey2009}
Ch. Bogey, N.~de~Cacqueray, and Ch. Bailly.
\newblock A shock-capturing methodology based on adaptative spatial filtering
  for high-order non-linear computations.
\newblock {\em J. Comp.\ Phys.}, 228(5):1447 -- 1465, 2009.

\bibitem{Carpenter1994220}
Mark~H. Carpenter, David Gottlieb, and Saul Abarbanel.
\newblock Time-stable boundary conditions for finite-difference schemes solving
  hyperbolic systems: Methodology and application to high-order compact
  schemes.
\newblock {\em Journal of Computational Physics}, 111(2):220 -- 236, 1994.

\bibitem{DucrosEtAl2000}
F.~Ducros, F.~Laporte, T.~Soul\`eres, V.~Guinot, P.~Moinat, and B.~Caruelle.
\newblock High-order fluxes for conservative skew-symmetric-like schemes in
  structured meshes: Application to compressible flows.
\newblock {\em Journal of Computational Physics}, 161(1):114 -- 139, 2000.

\bibitem{Gassner2013}
Gregor~J Gassner.
\newblock A skew-symmetric discontinuous galerkin spectral element
  discretization and its relation to sbp-sat finite difference methods.
\newblock {\em SIAM Journal on Scientific Computing}, 35(3):A1233--A1253, 2013.

\bibitem{Brouwer2013}
J\"orn~Sesterhenn Jens~Brouwer, Julius~Reiss.
\newblock Conservative time integrators of arbitrary order for skew-symmetric
  finite-difference discretizations of compressible flow.
\newblock {\em Computers \& Fluids}, submitted:xxx, 2013.

\bibitem{Kok2009}
J.C. Kok.
\newblock A high-order low-dispersion symmetry-preserving finite-volume method
  for compressible flow on curvilinear grids.
\newblock {\em Journal of Computational Physics}, 228(18):6811 -- 6832, 2009.

\bibitem{LaxLiu1998}
Peter~D. Lax and Xu-Dong Liu.
\newblock Solution of two-dimensional riemann problems of gas dynamics by
  positive schemes.
\newblock {\em SIAM J. Sci. Comput.}, 19(2):319--340, March 1998.

\bibitem{LeVeque1992}
Randall~J. LeVeque.
\newblock {\em Numerical methods for conservation laws}.
\newblock Lectures in Mathematics. Birkh\"auser-Verlag, Basel, 1992.

\bibitem{Morinishi2010}
Y.\ Morinishi.
\newblock Skew-symmetric form of convective terms and fully conservative finite
  difference schemes for variable density low-mach number flows.
\newblock {\em JCP}, 229(2):276--300, 2010.

\bibitem{Morinishi2013}
Y.~Morinishi and K.~Koga.
\newblock Skew-symmetric convection form and secondary conservative finite
  difference methods for moving grids.
\newblock {\em Journal of Comp. Phys.}, in press:xxx, 2013.

\bibitem{Morinishi1998}
Y.~Morinishi, T.S. Lund, O.V. Vasilyev, and P.~Moin.
\newblock Fully conservative higher order finite difference schemes for
  incompressible flow.
\newblock {\em Journal of Computational Physics}, 143(1):90 -- 124, 1998.

\bibitem{Morinishi2007}
Youhei Morinishi.
\newblock Forms of convecction and quadratic conservative finite difference
  schemes for low mach number compressible flow simulations.
\newblock {\em Trans. Jap. Soc. Mech. Engin. B}, pages 451--458, 2007.

\bibitem{Olsson1993}
Pelle Olsson.
\newblock Summation by parts, projections, and stability.
\newblock RICAS Tech.\ Rep.\ 93.04, 1993.

\bibitem{Pirozzoli2010}
Sergio Pirozzoli.
\newblock Generalized conservative approximations of split convective
  derivative operators.
\newblock {\em Journal of Computational Physics}, 229(19):7180 -- 7190, 2010.

\bibitem{Reiss2010}
Julius Reiss and Jörn Sesterhenn.
\newblock Conservative, skew–symmetric discretization of the compressible
  navier– stokes equations.
\newblock In Andreas Dillmann, Gerd Heller, Hans-Peter Kreplin, Wolfgang
  Nitsche, and Inken Peltzer, editors, {\em New Results in Numerical and
  Experimental Fluid Mechanics VIII}, volume 121 of {\em Notes on Numerical
  Fluid Mechanics and Multidisciplinary Design}, pages 395--402. Springer
  Berlin Heidelberg, 2013.

\bibitem{Strand1994}
Bo~Strand.
\newblock Summation by parts for finite difference approximations for d/dx.
\newblock {\em Journal of Computational Physics}, 110(1):47 -- 67, 1994.

\bibitem{Subbareddy2009}
Pramod~K. Subbareddy and Graham~V. Candler.
\newblock A fully discrete, kinetic energy consistent finite-volume scheme for
  compressible flows.
\newblock {\em Journal of Computational Physics}, 228(5):1347 -- 1364, 2009.

\bibitem{Thompson1985}
Joe~F. Thompson.
\newblock {\em Numerical Grid Generation}.
\newblock Elsevier, Amsterdam, 1985.

\bibitem{VerstappenVeldman2003}
R.~W. C.~P. Verstappen and A.~E.~P. Veldman.
\newblock Symmetry-preserving discretization of turbulent flow.
\newblock {\em JCP}, 187(1):343, 2003.

\end{thebibliography}






\end{document}